%% file: HR_V1.tex
\documentclass[11pt, a4paper]{article}
\usepackage[left=1.0in, right=1.0in, bottom=0.8in]{geometry}
\usepackage{iopams}
\expandafter\let\csname equation*\endcsname\relax
\expandafter\let\csname endequation*\endcsname\relax
\usepackage{amsmath,amssymb,amsfonts,amsthm}
\usepackage{psfrag,latexsym}
\usepackage{amssymb}
\usepackage{mathtools}
\usepackage{physics}
\usepackage{cancel} 
\usepackage{bm}
\usepackage{dcolumn}
\usepackage{epic} 
\usepackage{epsfig}
\usepackage{feynmp}
\usepackage{graphicx}
\usepackage{grffile}
\usepackage[usenames, dvipsnames]{xcolor} 
\usepackage[breaklinks,colorlinks = true,linkcolor = red,urlcolor=blue,citecolor=blue]{hyperref}
\usepackage{caption}
\usepackage{subcaption}
\usepackage{soul}
\usepackage{wrapfig}
\usepackage{xy} 
\usepackage{xcolor}
\usepackage{tikz}

\usepackage{makecell}

\usepackage{cite}
\usepackage{textcomp}
\usepackage{bbold}
\usepackage[scr=boondox,  
            cal=boondoxo]   
           {mathalpha}
\usepackage{ulem}
\usepackage[mathscr]{eucal}
\usepackage[ascii]{inputenc}
\input{macro}

\usepackage{etoolbox}
\makeatletter
\newrobustcmd{\fixappendix}{%
  \patchcmd{\l@section}{1.5em}{7em}{}{}%
  \patchcmd{\l@subsection}{2.3em}{7em}{}{}%
}
\makeatother

\usepackage{color}
\usepackage{dsfont}

\definecolor{dgreen}{rgb}{0,0.7,0}

\newcommand{\mc}[1]{\mathcal{#1}}  
\newcommand{\msc}[1]{\mathscr{#1}}  
\newcommand{\msf}[1]{\mathsf{#1}}  
  
\newcommand{\mf}[1]{\mathfrak{#1}} 
\newcommand{\mt}[1]{\mathtt{#1}}

\begin{document}

\title{Probing fluctuating hydrodynamics through current fluctuations in  hard rod gas}
\author{Tirthankar Mondal (tirthankar.mondal@icts.res.in) \\ International Centre for Theoretical Sciences, TIFR, Bengaluru -- 560089, India, \and Indranil Mukherjee (indranil.mukherjee@icts.res.in)\\ International Centre for Theoretical Sciences, TIFR, Bengaluru -- 560089, India \and Anupam Kundu (anupam.kundu@icts.res.in)\\ International Centre for Theoretical Sciences, TIFR, Bengaluru -- 560089, India}

\maketitle
\begin{abstract}
We study the full counting statistics of the integrated current in a one-dimensional hard-rod gas in equilibrium, employing two recently proposed fluctuating hydrodynamic frameworks: (i) Ballistic Macroscopic Fluctuation Theory (BMFT) and (ii) diffusive fluctuating hydrodynamic theory based on correlated Brownian motion approximation of hard-rod quasiparticles (CBFHD). We investigate how the predictions of these two approaches compare with each other and with the underlying microscopic dynamics. We show within BMFT that the current distribution admits a large-deviation form under ballistic scaling, and we derive an explicit expression for the large-deviation function, valid for an arbitrary inhomogeneous initial state factorized in point-particle coordinates. The corresponding current cumulants are also obtained. We subsequently analyze the current statistics within the CBFHD framework, which explicitly incorporates stochastic noise and dissipative terms, and demonstrate that the current distribution likewise obeys a large-deviation principle in this approach. We find that, while  BMFT predictions for the first few cumulants and the rate function are in excellent agreement with direct numerical simulations of the microscopic hard-rod dynamics, CBFHD prediction is not. We further verify the predictions of the CBFHD  independently using a collection of correlated Brownian motions, which provides an approximate description of the dynamics of hard-rod quasiparticles. Our findings indicate that CBFHD does not accurately capture large-scale current fluctuations in the hard-rod gas, whereas BMFT yields predictions in quantitative agreement with the microscopic dynamics, at least in the context of full counting statistics.
\end{abstract}

\newpage
\tableofcontents

\markboth{\it   }{}

\section{Introduction}
Understanding the collective behavior of many-body systems from their microscopic dynamics is one of the central goals of statistical physics. While equilibrium statistical mechanics provides a well-established framework for describing stationary properties, a wide range of physical phenomena occur out of equilibrium, where fluctuations of macroscopic observables play an essential role. In recent years, considerable attention has therefore been devoted not only to the evolution of average densities and currents, but also to their fluctuations and the emergence of universal large-scale dynamical behavior. 

Hydrodynamic theories provide a powerful framework for describing the long-time behavior of many body systems \cite{Forster1975, Spohn1991}. Conventional hydrodynamics describes systems with a finite number of conserved quantities and provides an effective description of the evolution of their local densities and currents. Over the past decades, this framework has been extended to integrable systems \cite{rigol2007relaxation,Pozsgay_2013,langen2015experimental, vidmar2016generalized}, which possess an extensive number of conserved quantities. This framework is known as the generalized hydrodynamics (GHD) \cite{castro2016emergent, bertini2016transport, doyon2019generalized, doyonlecturenotes, bastianello2022introduction, doyon_PRX_2025_GHD}. The GHD equations take the form of collisionless Boltzmann equations written in terms of the phase-space densities of stable quasiparticles \cite{doyonlecturenotes}, and these equations describe the evolution of conserved densities at the largest space-time scales, known as
the Euler scaling limit (in which space and time are related ballistically). To capture dynamics at smaller space-time scales, one must incorporate Navier-Stokes (NS)-type or more general diffusive corrections to GHD \cite{doyon2017dynamics, Hubner2025_PRL}. Interestingly, due to long-range correlations present in the Euler limit, integrable systems exhibit unusual diffusive corrections not captured by standard NS corrections \cite{doyon2023emergence,Hubner2025_SciPost}.
Although the Euler GHD, together with its diffusive corrections, has provided a successful description of the evolution of mean density and current profiles, the fluctuations of these observables at macroscopic scales are considerably less understood and have recently drawn a lot of interest both theoretically and experimentally. This motivates extending this hydrodynamic description beyond average conserved densities and currents to develop a statistical theory of fluctuations on macroscopic scales.

A classical framework for describing fluctuations at hydrodynamic scales was developed by Landau and Lifshitz \cite{landau2013statistical,landau1992hydrodynamic}. In this framework, fluctuations are incorporated through stochastic contributions to the dissipative currents, leading, within the linear-response regime, to linear fluctuating hydrodynamic equations \cite{Zarate2006hydrodynamic,de2006hydrodynamic}. Such a linearized hydrodynamic (HD) description, however, is insufficient to capture transport phenomena in certain systems, notably anomalous transport in low-dimensional systems \cite{alder1970decay,forster1977large,van2012exact}, where an extension to nonlinear fluctuating hydrodynamics is required \cite{spohn2014nonlinear,spohn2016fluctuating}. For systems exhibiting diffusive transport, macroscopic fluctuation theory (MFT) provides a systematic framework based on large-deviation principles \cite{bertini2002macroscopic,bertini2015macroscopic}, characterizing the probabilities of density and current fluctuations on diffusive space-time scales.

More recently, a related large-deviation framework, known as ballistic macroscopic fluctuation theory (BMFT), has been developed for integrable systems to describe density and current fluctuations on ballistic scales \cite{doyon2023emergence,doyon2023ballistic,Kundu_2025}. The theory is based on the principle of local relaxation, according to which fluctuations of local observables at the largest space-time scales are determined by fluctuations of the conserved densities, coherently propagated from the initial state. In the hydrodynamic projection of coarse-grained observables onto the space of conserved quantities, the projected-out component generally gives rise to hydrodynamic noise and ``bare'' diffusion coefficients, related through an Einstein relation. However, it has recently been shown that hydrodynamic noise, and consequently bare diffusion, must vanish in integrable systems \cite{doyon2025hydro}. It was further argued that BMFT then provides the appropriate description of hydrodynamic fluctuations in integrable models to all orders in the inverse variation scale of connected correlation functions \cite{doyon2025hydro}.

In contrast, a fluctuating hydrodynamic theory has recently been developed for a homogeneous hard-rod gas in equilibrium, based on  drifted and correlated Brownian motion approximation for the quasiparticle dynamics\cite{Chahal_tracer_2026}. We refer to this as CBFHD, which explicitly incorporates both stochastic noise and dissipative terms. Since the hard-rod gas is an integrable system, this formulation appears, at first sight, to be at variance with the picture described above. This naturally raises two questions: Are the two fluctuating hydrodynamic descriptions equivalent? How do their predictions for macroscopic fluctuations compare with each other and with those obtained directly from the underlying microscopic dynamics? While addressing the first question appears to be challenging, in this work we focus on the second by considering the full counting statistics of the integrated current in a one-dimensional hard-rod system in homogeneous equilibrium. We apply both hydrodynamic formulations to this problem and systematically compare their predictions with results obtained from direct numerical simulations of the microscopic dynamics.

The hard-rod gas is a paradigmatic integrable many-body system composed of extended particles of length $a$ and unit mass, interacting through hard-core repulsion. The collisions are elastic, and the particles exchange momenta upon each collision. Originally introduced by Tonks in 1930 as one of the simplest models of an interacting gas \cite{tonks1936HR}, the hard-rod system has since served as a useful setting for studying a variety of thermodynamic and statistical properties, including density fluctuations, pair-correlation functions, and distribution functions \cite{robledo1986distribution,salsburg1953molecular,sells1953pair,koppel1963partition}. At the macroscopic scale, the hydrodynamic equation for the phase-space density of the hard-rod gas was first formulated by Percus based on the Boltzmann equation \cite{percus1969exact}. With the development of GHD, the hard-rod gas has emerged as an important testbed for examining the predictions of both GHD and BMFT for macroscopic observables. In particular, several key features of these hydrodynamic frameworks have been demonstrated explicitly through performing both the microscopic and macroscopic analyses \cite{doyon2017dynamics,singh2024thermalization,powdel2024conserved,Mukherjee2026correlation}.

The integrated current provides a natural observable for characterizing transport fluctuations, as it directly quantifies the cumulative transport of particles across a fixed spatial location. For the hard-rod system, the integrated current $\mc Q(\msf T)$, across the origin over a time interval $[0,\msf T]$, can be interpreted as the net change in the number of rods to the right of the origin between the initial time and the time $\msf T$. We investigate its statistics at both microscopic and hydrodynamic scales, by exploiting the exact mapping between the dynamics of hard rods and those of hard-point particles. This exact mapping, which has also been employed in previous studies \cite{lebowitz1968time,derrida2007non,bernstein1988expansion}, enables us to obtain the full distribution of the integrated current in hard-rod system. The microscopic description, obtained in this work, provides an exact benchmark against which hydrodynamic predictions can be tested.

For diffusive systems, MFT has provided a powerful framework for describing hydrodynamic evolution and characterizing transport fluctuations, spatial correlations, and the full counting statistics (FCS) of currents \cite{2010_Lecomte_Current,2014_Krapivsky_Large,2015_Krapivsky_Tagged,2016_Sadhu_Correlations,2019_Derrida_Large,Bernard2021CanMFTBeQuantized,Sadhu2018LargeDevTracer,2023_Agranov_Macroscopic,Mallick2024ExactMFTIntegrable,Dandekar_2024,Bodineau_2025,saha_LD_PRE2026,jangid2026MFT}. For systems exhibiting ballistic transport, however, a corresponding description remains considerably less explored. Recent studies have employed BMFT to investigate the FCS of integrated currents in homogeneous hard-rod gas \cite{Jitendra_BMFT,saha2026universaltracerstatisticssinglefile}. In the present work, we exploit the exact mapping between hard rods and hard-point particles to map the current FCS of the hard-rod gas onto corresponding problems for point particles, with appropriate tracer-particle problems appearing as intermediate steps. This mapping allows us to obtain the full distribution of the integrated current directly from the microscopic dynamics. To obtain the large time behavior, we apply BMFT and explicitly show that the current distribution admits a large-deviation form with ballistic scaling. The above mentioned mapping and the BMFT  approach allowed us to determine the corresponding rate function explicitly for a class of inhomogeneous initial states. Our work thereby provides an explicit expression that was not obtained in the previous studies.

We next turn to the CBFHD approach. Following this formulation, we perform the calculation directly in the hard-rod coordinates and determine the statistics of the integrated current. Within CBFHD, we again find a large-deviation form with ballistic scaling, as in the BMFT description. Interestingly, although the two hydrodynamic approaches predict the same ballistic scaling, they yield different rate functions. Comparing these predictions with direct numerical simulations of the microscopic hard-rod dynamics, we find agreement with the BMFT rate function, whereas the CBFHD prediction deviates from the microscopic results. This comparison with the underlying microscopic dynamics thus provides a direct assessment of the predictions of the two hydrodynamic formulations for macroscopic current fluctuations in the hard-rod gas.

The paper is organized as follows. (i) Sec. \ref{sec:2}  introduces the microscopic dynamics of the hard-rod system and the mapping to the corresponding hard-point gas. We then specify the initial condition considered in this work. We end this section by briefly discussing the two hydrodynamic approaches, BMFT and CBFHD, used to characterize the statistics of the integrated current at the macroscopic scale. 
(ii) In Sec. \ref{sec:BMFT}, we study the integrated-current statistics within BMFT framework. The BMFT equations are derived using a saddle-point approach, and the full current distribution is obtained. We show explicitly that the distribution exhibits a large-deviation form and determine the corresponding rate function. (iii) Sec. \ref{CBFHD} discusses the CBFHD formulation. Using a saddle-point calculation, we determine the full statistics of the integrated current, including its large-deviation behavior. This section ends with the comparison of these two hydrodynamic descriptions with the exact microscopic dynamics. (iv) Finally, Sec. \ref{sec:conclusion} summarizes our main results and conclusions.  A derivation of the integrated-current distribution using the microscopic approach and other supplementary calculations are provided in the appendix.

\section{Hard rods system and integrated current}\label{sec:2}
We consider a one-dimensional system of $N$ identical hard rods, each with unit mass and length $a$, moving on an infinitely extended line. The position and velocity of the rods are represented by $\{X_j,V_j\}$, for $j=-N_1,...,0,...,N_2-1$ with $N=N_1+N_2$. Throughout this work, we adopt the following labeling convention. The rods with position greater or equal to the origin are labeled as $j=0,1,2,...$ according to the increasing order of their distance from the origin. Similarly the rods with position strictly less than the origin are labeled as $j=-1,-2,-3,...$ again according to the increasing order of their distance from the origin.  
Thus, the labels increase sequentially from left to right, with negative indices denoting rods to the left of the origin and positive indices denoting rods to its right such that $j=- N_1,\dots,-1,0,1,\dots, N_2-1$. Consequently, the positions of the rods are ordered such that $X_j<X_{j+1}-a$. Since the mass is set to unity, the momentum of each rod coincides with its velocity. The rods undergo free ballistic motion between instantaneous elastic collisions. At each collision the velocity between the two colliding rod are exchanged. In the point-particle limit $a\to0$, the hard-rod system reduces to a gas of hard-point particles (HPG). In this limit, the Jepsen mapping \cite{jepsen1965dynamics} allows the interacting hard-point system to be represented in terms of non-interacting particles. More generally, the dynamics of hard rods can be related exactly to those of hard-point particles through a coordinate transformation \cite{percus1969exact,bernstein1988expansion,lebowitz1968time}. Specifically, given a hard-rod configuration $\{X_j,V_j\}$, we introduce a corresponding set of point-particle coordinates $\{x_j,v_j\}$ according to
\begin{equation}\label{eq:HP_mapping}
x_j = X_j-\left(j+\frac{1}{2}\right)a,\qquad v_j=V_j,~~~\text{for}~~ j=- N_1,\dots,0,\dots,  N_2-1. 
\end{equation}
The above transformation is essentially obtained by excluding the inaccessible space between successive rods.
In this point-particle representation, the dynamics are greatly simplified: unlike hard rods, they exchange their velocities during collisions without experiencing any jump in their positions, i.e., there is no positional jump.  Instead of exchanging the velocities, one can exchange their labels to map the dynamics of the hard-point gas to a gas of non-interacting particles \cite{Chahal_tracer_2026}. Since
the transformation in Eq.~\eqref{eq:HP_mapping} is one-to-one, the hard-rod dynamics are exactly solvable by mapping the system back to hard-rod coordinates from the corresponding point-particle coordinates using the inverse of the transformation in Eq.~\eqref{eq:HP_mapping}.

In this paper, we are interested in studying the statistics of the integrated current $\mc Q(\msf T)$ across the origin in a one-dimensional hard-rod system. In terms of the single-particle phase-space density $\mf f(X,v,t)$, the net integrated current up to time $\msf T$ can be expressed as
\begin{align}
\label{def:Q(T)}
 \mc Q(\msf T)=\int_0^{\infty}dX\int_{-\infty}^{\infty} dv~[\mf f(X,v,\msf T)-\mf f(X,v,0)].
\end{align}
Thus, $\mc Q(\msf T)$ represents the difference in the number of rods to the right of the origin between time $\msf T$ and the initial time.  
We investigate the probability distribution $\mc P_{\rm r}(\mc Q,\msf T)$ of the integrated current and its associated cumulants using a microscopic approach and two hydrodynamic descriptions of large-scale fluctuations: (i) BMFT and (ii) CBFHD. For the microscopic approach and the BMFT description, the current distribution $\mc P_{\rm r}(\mc Q,\msf T)$ is obtained by exploiting the exact mapping between hard rods and hard-point particles.
The mapping can be understood as follows: We consider the hard-rod system in which the positions of the rods are labeled as $\{X_j\}$, where the rods initially to the right of the origin carry the labels $j\geq0$, and those to the left carry negative labels. For $j>0$, the condition $\mc Q(\msf T)>j$ implies that the rod initially labeled as $-j$ has reached the right to the origin by time $\msf T$, {\it i.e.}, $X_{-j}(\msf T)>0$. Here the rod labeled $-j$ denotes the $j$-th rod counted from the right among those initially located to the left of the origin, and $X_{-j}(\msf T)$ is its position at time $\msf T$. 
Consequently, the probability $P^{\rm r}(X_{-j}(\msf T)>0)$ that this rod lies to the right of the origin at time $\msf T$ equals the cumulative probability that the integrated current exceeds $j$, {\it i.e.}, $\mathbb{P}^c_{\rm r}(\mc Q(\msf T)>j)=P^{\rm r}(X_{-j}(\msf T)>0)$; where $\mathbb{P}^c_{\rm r}(j)$ denotes the cumulative probability in the hard-rod system. Using the exact mapping between the hard rod and hard point particles given in Eq.~\eqref{eq:HP_mapping}, the condition  $X_{-j}(\msf T)>0$ translates into the position of the corresponding point particle labeled $-j$, as $x_{-j}(\msf T)>(j-1/2)a$.  This implies $P^{\rm p}(x_{-j}(\msf T)>(j-1/2)a)=P^{\rm r}(X_{-j}(\msf T)>0)$, where $P^{\rm p}(x_{-j}(\msf T)>z)$ denotes the probability that the point particle $-j$ lies right to the position $z$ at time $\msf T$. Finally, this probability $P^{\rm p}(x_{-j}(\msf T)>(j-1/2)a)$ can then be related to the cumulative probability in the hard-point particle system. Specifically, $P^{\rm p}(x_{-j}(\msf T)>(j-1/2)a)=\mathbb{P}^c_{\rm p}(\mc Q^{\rm p}_{\frac{(j-1/2)a}{\msf T}}(\msf T)>j)$, where $\mc Q^{\rm p}_{u}(\msf T)$ represents the net number of point particles crossing the spacetime line $x(t)=ut$ for $u\ge 0$ from {\it left to right} at time $\msf T$. Combining these relations, we obtain the equivalences
\begin{align}\label{eq:PP_Rod_equivalences}
   \mathbb{P}^c_{\rm r}(\mc Q(\msf T)>j)=P^{\rm r}(X_{-j}(\msf T)>0)= P^{\rm p}(x_{-j}(\msf T)>(j-1/2)a)=\mathbb{P}^c_{\rm p}\left(\mc Q^{\rm p}_{\frac{(j-1/2)a}{\msf T}}(\msf T)>j\right).
\end{align}
By computing any one of the above probabilities, one can obtain the current distribution $\mc P_{\rm r}(\mc Q,\msf T)$ for $\mc Q >0$. To obtain the distribution for $\mc Q <0$, an  analogous argument can be constructed to get another equivalence relation
\begin{align}\label{eq:PP_Rod_equivalences_j_-ve}
   \mathbb{P}^c_{\rm r}(\mc Q(\msf T)<-j)=P^{\rm r}(X_{j-1}(\msf T)<0)= P^{\rm p}(x_{j-1}(\msf T)<-(j-1/2)a)=\mathbb{P}^c_{\rm p}\left(\mc Q^{\rm p}_{-\frac{(j-1/2)a}{\msf T}}(\msf T)>j\right),
\end{align}
for $j\ge0$, where $\mc Q^{\rm p}_{-u}(\msf T)$ represents the net number of point particles crossing the spacetime line $x(t)=-ut$ for $u\ge 0$ from {\it right to left} at time $\msf T$.

The distribution of $\mc P_{\rm r}(\mc Q=j,\msf T)$, for for $j>0$ can then be obtained using the point particle problem in Eq.~\eqref{eq:PP_Rod_equivalences} as
\begin{align}\label{eq:PP_Rod_dist_relation}
\mc P_{\rm r}(\mc Q=j,\msf T)&=P^{\rm p}(x_{-j}\left(\msf T)>(j-1/2)a\right)-P^{\rm p}\left(x_{-(j+1)}(\msf T)>(j+1/2)a\right)\cr
&=\mathbb{P}^c_{\rm p}\left(\mc Q^{\rm p}_{\frac{(j-1/2)a}{\msf T}}(\msf T)>j\right)-\mathbb{P}^c_{\rm p}\left(\mc Q^{\rm p}_{\frac{(j+1/2)a}{\msf T}}(\msf T)>j+1\right).
\end{align}
The corresponding distribution for $\mc Q<0$ follows analogously from Eq.~\eqref{eq:PP_Rod_equivalences_j_-ve}.

In this paper, to compute this distribution of net integrated current across the origin following  the microscopic dynamics and BMFT, we consider a class of initial configurations that are specified first in the hard-point representation and subsequently mapped onto the corresponding hard-rod coordinates. The initial positions and velocities of $N$ point particles, denoted by ${x_j',v_j}$, are sampled independently from the probability density $\Omega(x',v)$ over the interval $[-L_1,L_2] \times [-\infty,\infty]$, with the normalization condition $\int_{-L_1}^{L_2} dx\int dv~\Omega(x,v)=1$. In particular we assume that the initial state is inhomogeneous over a length scale $\ell$ {\it i.e.,} $\Omega(x,v)=\Psi(x/\ell,v)$. The sampled positions are then arranged in ascending order according to $\{x_j\}=$ Order[$\{x_j'\}$] with $-L_1<x_{-N_1}<...<x_{ N_2-1}<L_2$. The marginal velocity distribution of individual particles is $\mc g(v) = \int_{-L_1}^{L_2}dx~\Omega(x,v)$. The positions and velocities of $N$ particles are chosen from  the following  joint distribution 
\begin{align}
\mathfrak{P}(\{x_{j},v_{j}\},0) = \sum_{N_1=0}^N\binom{N}{N_1} \mc q_{l}^{N_1} \mc q_r^{N-N_1}~\times~\mathfrak{P}_{N_1}(\{x_{j},v_{j}\},0), 
\label{eq:prob_init}
\end{align}
where 
$\mc q_{l} = \int_{-L_1}^0dx \int dv~ \Omega(x,v)$ and $\mc q_r = \int_0^{L_2}dx \int dv ~\Omega(x,v)$ and the conditional joint distribution $ \mathfrak{P}_{N_1}(\{x_{j},v_{j}\},0)$ is provided in Appendix~\ref{app:P_ini} in detail.

The mean phase-space density (PSD) and the mass density of the hard-point particles corresponding to the above initial distribution, are given, respectively, by $f_{\rm leq}(x,v)=N\Omega(x,v)$ and $\varphi(x)=N\int dv~\Omega(x,v)$. We subsequently consider the thermodynamic limit in which both the number of particles, $N\to \infty$, and $|L_2-L_1|\to \infty$, while the initial mass-density profile $\varphi(x)$ is kept finite at every position. For any configuration of point particles ${x_j,v_j}$, the corresponding hard-rod configuration is obtained by applying the inverse of the mapping in Eq.~\eqref{eq:HP_mapping}. In particular, the rod positions  are given by
\begin{align}
X_j=x_j+\left(j+\frac{1}{2}\right)a,
\qquad
j=-N_1,\ldots,0, \ldots,N_2-1.
\label{eq:map_pp_rod}
\end{align}
Since velocities do not change under transformation, we, from now on,  denote the velocities of the rods and the corresponding point particles by $\{v_j\}$. 
In the thermodynamic limit, we have an ensemble of hard-rod configurations characterized by the average mass density profile $\varrho(X(x)) = \frac{\varphi(x)}{1+a\varphi(x)}$, where one can choose the following transformation
\begin{align}
X(x)=x+a\int^x_{-\infty} dy~\varphi(y), \label{transf-x->X}    
\end{align} 
between the point particle coordinate $x$ and the corresponding hard-rod coordinate $X(x)$.

To study the statistics of the integrated current $\mc Q(\msf T)$ at the macroscopic scale, we employ two hydrodynamic approaches: BMFT and CBFHD. The hydrodynamic description of the hard-rod system can be described in terms of the single-particle phase-space density $\mf f(X,v,t)$, which evolves according to the Euler-scale generalized hydrodynamics (GHD) equation on ballistic space-time scales, as \cite{powdel2024conserved, singh2024thermalization}
\begin{align} \label{eq:Euler_GHD}
\begin{split}
&\partial_t \mf f(X,v,t) +\partial_X \big(\msf V^{\rm eff}(X,v,t) \mf f(X,v,t)\big) = 0,\cr 
&{\rm with}~~ \msf V^{\rm eff}(X,v,t) = \frac{v-a\int dv~v~\mf f ( X,v,t)}{1-a \int dv~ \mf f( X,v,t)}
\end{split}
\end{align}
According to BMFT, fluctuations in the phase space density on the Euler space-time scale are dominated by the fluctuations in the initial conditions, which are coherently propagated to the current location by the Euler GHD Eq.~\eqref{eq:Euler_GHD}. In analogy with macroscopic fluctuation theory (MFT), one can formulate a large deviation principle to describe fluctuations in the phase space density at any later time, and this formulation can completely describe fluctuations in macroscopic observables. 

Recently, Ferrari and Olla proposed a fluctuating hydrodynamic theory for a gas of hard rods in equilibrium \cite{ferrari2023macroscopic}, characterized by a homogeneous mass density $\varrho_0$ and a velocity distribution $\mc g(v)$, which explicitly incorporates the noise and dissipation terms. It is known that the hard-rod gas can be viewed equivalently as a gas of quasiparticles (which are obtained by exchanging the labels of the two rods at each  collision instead of momenta) \cite{Chahal_tracer_2026}. According to the Ferrari-Olla theory, these quasiparticles effectively undergo Brownian motion with an effective drift, however the noises acting on different quasiparticles are correlated. The effective equation of motion for a quasiparticle with bare velocity $v$ is given by
\begin{subequations}
\label{eq:corr-BM}
\begin{align}
    \frac{dX_v}{dt}= \msf V^{\rm eff}(v) +\eta_v(t),~~\text{with}~~\msf V^{\rm eff}(v)=\frac{v}{1-a\varrho_0}, \label{Brownian-QP}
\end{align}
The stochastic variables $\eta_v(t)$ are mean-zero white Gaussian noises with covariance
\begin{align}\label{eq:noise_corr}
\begin{split}
    \langle \eta_v(t)\eta_u(t')\rangle &= \frac{\mc{G}(v,u) }{\sqrt{\mc{D}(v)}\sqrt{\mc{D}(u)}}~\delta(t-t'),
    \end{split}
\end{align}

\begin{align}\label{eq:D_v}
\text{where}~~~~~~~~~~~~~~~~
    \mathcal{D}(v)&= a^2 \varphi_0 \int dw |v-w| \mc{g}(w), ~~~~~~~~~~~
    \text{and, } \\
    \mc{G}(u,v) &=\frac{1}{2}\left( \mc{D}(v)+\mc{D}(u)- a^2 \varphi_0|v-u|\right), \label{eq:G_uv}
\end{align}
with $\varphi_0=\frac{\varrho_0}{1-a\varrho_0}$.
\end{subequations} 
Recently, starting from the effective microscopic Brownian dynamics of the quasiparticles in Eq.~\eqref{Brownian-QP}, an equivalent fluctuating hydrodynamic description has been derived  in the original coordinates using a heuristic formalism \cite{Chahal_tracer_2026}. The resulting equation reads
\begin{align}
\partial_t \mf f(X,v,t)+\msf V^{\rm eff}(v)\partial_X \mf f(X,v,t)=\frac{\mc D(v)}{2}\partial_X^2 \mf f(X,v,t) +\partial_X \{\mf f(X,v,t)\sqrt{\mc D(v)} \xi(v,t)\}. \label{eq:FOFHD_eq1}
\end{align}
where  $\xi(v,t)$ is a zero-mean white Gaussian noise field with the same correlation as in Eq.~\eqref{eq:noise_corr}. 
After recentering the position coordinate  on  Euler characteristics and then considering the  diffusive space-time scaling, the FHD satisfied by the phase space density looks similar to the following fluctuating hydrodynamics derived by 
Ferrari and Olla  \cite{ferrari2023macroscopic}.
The statistics of the integrated current $\mc Q(\msf T)$ at the macroscopic scale are investigated in the following sections using these two formulations of fluctuating hydrodynamics -- BMFT and CBFHD, separately. The computation of the distribution of $\mc Q(\msf T)$ using  microscopic approach is presented in Appendix~\ref{sec:app_micro}.

\section{Integrated-current statistics using BMFT}\label{sec:BMFT}
This section aims to obtain the  statistics of $\mc Q(\msf T)$, including its large-deviation within the BMFT framework. 
Recently this framework  has been employed  to obtain the full counting statistics in a homogeneous state \cite{Jitendra_BMFT, saha2026universaltracerstatisticssinglefile}, where it was shown that the distribution of the current fluctuations admits a large deviation form under ballistic scaling however the expression of the rate function was not obtained explicitly. In this work, exploiting the exact mapping between hard rods and hard-point particles within the BMFT framework, we derive an explicit expression of the large deviation function, valid for arbitrary inhomogeneous initial state factorized in point particle coordinates (see Eq.~\eqref{eq:prob_init}).

As established by the equivalence relation in Eq.~\eqref{eq:PP_Rod_equivalences}, the hard-rod current statistics can be related directly to those of the corresponding hard-point-particles. 
In particular, the mapping reduces the determination of the hard-rod current distribution $\mc P_{\rm r}(\mc Q,\msf T)$ to two problems of current distribution in point-particle system. More precisely, to find the distribution $\mc P_{\rm r}(\mc Q,\msf T)$  for $\mc Q \ge 0$, we solve the distribution of point-particle current $\mc Q^{\rm p}_u(\msf T)$ across the spacetime line $x(t)=ut$ with $u\ge 0$ from {\it left to right}. Similarly, to obtain $\mc P_{\rm r}(\mc Q,\msf T)$  for $\mc Q < 0$ we solve the distribution of current $\mc Q^{\rm p}_{-u}(\msf T)$ flowing across the spacetime line $x(t)=-ut$, with $u\ge 0$ from {\it right to left}. 

To proceed we first transform the HD description of hard-rods in Eq.~\eqref{eq:Euler_GHD} to HD description of point particles by transforming hard-rod coordinates to point particles following 
\begin{align}
x(X)=X-a\int^X_{-\infty} dY\int_{-\infty}^\infty dv~\mf f(Y,v,t). \label{transf-X->x}    
\end{align}
This transformation is, in fact, the inverse of the transformtion in Eq.~\eqref{transf-x->X}. Under this transformation the hard-rod density $\mf f(X,v,t)$ transform to point-particle density function $f(x,v,t)$
\cite{singh2024thermalization, powdel2024conserved, Kundu_2025} as
\begin{align}
f(x(X),v,t) = \frac{\mf f(X,v,t)}{1-a \int_{-\infty}^\infty dw ~\mf f(X,w,t)}.
\end{align}
Consequently the HD Eq.~\eqref{eq:Euler_GHD} transforms to the following Euler equation of free point particles
\begin{align}\label{eq:PP_Euler}
    \partial_tf(x,v,t)+v\partial_xf(x,v,t)=0,
\end{align}
valid over the region $-\infty \le x \le \infty$ and $-\infty \le v \le \infty$.

The net integrated particle current $\mc Q^{\rm p}_{u}(\msf T)$ crossing across the line $x(t)=ut$ over duration $[0, \msf T]$ can now be expressed easily in terms of the single-particle phase-space density of point particles, $f(x,v,t)$, as
\begin{align}\label{eq:PP_QT}
    \mc Q^{\rm p}_{u}(\msf T)=\int dx\int dv~\Big[\Theta(x-u\msf T)f(x,v,\msf T)-\Theta(x)f(x,v,0)\Big].
\end{align}
Analogously, the integrated particle current $Q^{\rm p}_{-u}(\msf T)$ crossing the line $x(t)=-ut$ from {\it right to left} over time  $[0, \msf T]$ is given by 
\begin{align}\label{eq:PP_QT_u<0}
    \mc Q^{\rm p}_{-u}(\msf T)=\int dx\int dv~\Big[\Theta(-x-u\msf T)f(x,v,\msf T)-\Theta(-x)f(x,v,0)\Big].
\end{align}
For a given initial phase-space profile $f(x,v,0)$, the solution of Eq.~\eqref{eq:PP_Euler} is uniquely determined, and consequently the observables $\mc Q^{\rm p}_{u}(\msf T)$ and $\mc Q^{\rm p}_{-u}(\msf T)$, defined in Eqs.~\eqref{eq:PP_QT}-\eqref{eq:PP_QT_u<0}, are deterministic.  Its randomness solely originates from the statistical fluctuations of the initial configuration $f(x,v,0)$. For  initial configurations chosen from the distribution in Eq.~\eqref{eq:prob_init}, it can be shown that in the thermodynamic limit, the probability of observing
the initial phase-space density $f(x,v,0)$ (denoted by $\msc{P}[f(x,v,0)]$) is described by the large deviation principle characterized by the free energy functional \cite{Kundu_2025}
\begin{align}\label{eq:PP_freeenergy_F_f}
    \mathcal{F}^{\rm p}[f(0)]=\int dx \int dv ~f(x,v,0)~[\psi(x,v)+\mu+\ln f(x,v,0)],
\end{align}
with $\psi(x,v)=-\ln\Omega(x,v)$. More precisely, the probability  is  given by
\begin{align}
    \msc{P}[f(x,v,0)]=\int d\mu ~{\rm e}^{\mu N} ~{\rm e}^{-\Delta\mathcal{F}^{\rm p}[f(x,v,0)]}, \label{eq:mscP[f(0)]}
\end{align}
where $\Delta \mathcal{F}^{\rm p}[f]=\mathcal{F}^{\rm p}[f]-\mathcal{F}^{\rm p}[f_{\rm leq}]$  and $\mu$ is a Lagrange multiplier (or chemical potential) that ensures the normalization of the mass $\int dx\int dv~ f(x,v,0)=N$. The density 
$f_{\rm leq}(x,v)$ denotes the local equilibrium phase space density of the hard point particles that minimizes the free energy functional $\mathcal{F}^{\rm p}[f(0)]$ satisfying $\int dx\int dv~ f_{\rm leq}(x,v)=N$. In fact, one finds 
\begin{align}
f_{\rm leq}(x,v) \propto e^{-\psi(x,v)} = \Omega(x,v).
\label{eq:f_leq}
\end{align}
The free energy difference in Eq.~\eqref{eq:mscP[f(0)]} can be expressed in terms of $f_{\rm leq}(x,v)$ as \cite{Kundu_2025}
\begin{align}\label{eq:e^(-DelF)-1}
    \Delta \mathcal{F}^{\rm p}[ f(0), f_{\rm leq}]&=\int dx\int dv\Bigg[  f(x,v,0)\ln\Big(\frac{ f(x,v,0)}{ f_{\rm leq}(x,v)}\Big)-\big( f(x,v,0)- f_{\rm leq}(x,v)\big)\Bigg].
\end{align}


\noindent
Once the initial density profile $f(x,v,0)$ is chosen, it evolves accoding to the Euler equation \eqref{eq:PP_Euler}. The probability distribution of the time evolution history of the profile $f(x,v,t)$ over any time duration $[0,\msf T']$ can now be written as   \cite{Kundu_2025}
\begin{align}\label{eq:PP_Pfxvt}
    \msc{P}[\{f(x,v,t);0\leq t \leq \msf T'\}]\asymp \int d\mu ~{\rm e}^{\mu N}~{\rm e}^{-\Delta\mathcal{F}^{\rm p}[f(x,v,0)]} ~\delta \Big[\partial_tf(x,v,t)+v\partial_xf(x,v,t)\Big].
\end{align}
To understand the statistical properties of $\mc Q^{\rm p}_{u}(\msf T)$ and $\mc Q^{\rm p}_{-u}(\msf T)$,  it is convenient to consider their generating functions, defined as
\begin{align}\label{eq:PP_MGF1}
    \big\langle {\rm e}^{-\lambda \mc Q^{\rm p}_{u}(\msf T)}\big\rangle&=\int \mathcal{D}[f(x,v,t)]~\msc{P}[\{f(x,v,t);0\leq t \leq \msf T\}]~{\rm e}^{-\lambda \mc Q^{\rm p}_{u}(\msf T)}, \cr 
    \big\langle {\rm e}^{-\lambda \mc Q^{\rm p}_{-u}(\msf T)}\big\rangle&=\int \mathcal{D}[f(x,v,t)]~\msc{P}[\{f(x,v,t);0\leq t \leq \msf T\}]~{\rm e}^{-\lambda \mc Q^{\rm p}_{-u}(\msf T)} .
\end{align}
In the following we focus on the explicit computation of $\big\langle {\rm e}^{-\lambda \mc Q^{\rm p}_{u}(\msf T)}\big\rangle$; the generating function $\big\langle {\rm e}^{-\lambda \mc Q^{\rm p}_{-u}(\msf T)}\big\rangle$ can be obtained by an analogous procedure. Substituting the expression of $\msc{P}[\{f(x,v,t);0\leq t \leq \msf T\}]$ from Eq.~\eqref{eq:PP_Pfxvt} into Eq.~\eqref{eq:PP_MGF1} and using the integral representation of the delta functional through an auxiliary field $h(x, v, t)$, Eq.~\eqref{eq:PP_MGF1} can be expressed as
\begin{align}\label{eq:GF_BMFT}
   \big\langle e^{-\lambda \mc Q_u^{\rm p}(\msf T)}\big\rangle=\int d\mu ~{\rm e}^{\mu N}\int \mathcal{D}[f(x,v,t), h(x,v,t)]~{\rm e}^{-\mathcal{S}_{\lambda}[f(x,v,t),h(x,v,t)]}
\end{align}
and the action $\mathcal{S}_{\lambda}[f,h]$ is given by
\begin{align}\label{eq:PP_BMFT_action}
    \mathcal{S}_{\lambda}[f,h]=&\Delta \mathcal{F}^{\rm p}[f(0), f_{\rm leq}]+\int_0^{\msf T} dt\int dx \int dv~h(x,v,t)~\Big[\partial_tf(x,v,t)+v\partial_xf(x,v,t)\Big]\cr
    &~~~~~~~~~~~~~~~~~+\lambda \int dx \int dv ~\Big[\Theta(x-u\msf T)f(x,v,\msf T)-\Theta(x)f(x,v,0)\Big]
\end{align}
where we have used the definition of $\mc Q^{\rm p}_{u}(\msf T)$ as given in Eq.~\eqref{eq:PP_QT}.
As we are interested in the fluctuations at length scale of $\mathcal{O}(\ell)$, we rescale the variables $\mc z=x/\ell$, $\tau=t/\ell$ and $\msc T=\msf T/\ell$. Under this rescaling, the phase space densities $f(x,v,t)$,  $f_{\rm leq}(x,v)$ and the auxiliary field $h(x, v, t)$ are expressed in terms of the rescaled variables, respectively, as  $f(x,v,t) \to \mc{f}(\mc z,v,\tau)$, $f_{\rm leq}(x,v) \to \mc{f}_{\rm leq}(\mc z,v)$, and $h(x,v,t) \to \mc{h}(\mc z,v,\tau)$. 
Consequently, the action can now be written as
\begin{subequations}\label{eq:PP_BMFT_action_scaled}
\begin{align}\label{eq:PP_BMFT_action_scaled1}
    \mathcal{S}_{\lambda}[f(x,v,t),h(x,v,t)] = \ell\msc{S}_{\lambda}[\mc{f}(\mc z,v,\tau),\mc h(\mc z,v,\tau)],
\end{align}
with
\begin{align}\label{eq:PP_BMFT_action_scaled2}
    \msc{S}_{\lambda}[\mc{f},\mc{h}]=&\msc{F}^{\rm p}[\mc{f}(0)]-\msc{F}^{\rm p}[\mc{f}_{\rm eq}]+\int_0^{\msc T} d\tau\int d\mc z \int dv~\mc{h}(\mc z,v,\tau)~\Big[\partial_{\tau}\mc{f}(\mc z,v,\tau)+v\partial_{\mc z}\mc{f}(\mc z,v,\tau)\Big]\cr
    &~~~~~~~~~~~~~~~~~+\lambda \int d\mc z \int dv ~\Big[\Theta(\mc z-u\msc T)\mc{f}(\mc z,v,\msc T)-\Theta(\mc z)\mc{f}(\mc z,v,0)\Big],
\end{align}
\end{subequations}
The rescaled free energy functional takes the form
\begin{align}\label{eq:PP_freeenergy_F_f_scaled}
    \Delta \msc{F}^{\rm p}[ \mc f(0),\mc f_{\rm leq}]&=\int dz\int dv\Bigg[  \mc f(z,v,0)\ln\Big(\frac{ \mc f(z,v,0)}{ \mc f_{\rm leq}(z,v)}\Big)-\big( \mc f(z,v,0)- \mc f_{\rm leq}(z,v)\big)\Bigg].
\end{align}
As the action in Eq.~\eqref{eq:PP_BMFT_action_scaled} scales linearly with $\ell$, the generating function in Eq.~\eqref{eq:GF_BMFT} can be evaluated in the large $\ell$ limit by a saddle-point method {\it i.e.,}
\begin{align}
 \big\langle e^{-\lambda \mc Q_u^{\rm p}(\msf T)}\big\rangle \asymp e^{-\ell \msc{S}_{\lambda}[\mc f^{\rm sd}_\lambda(\mc z,v,\tau),\mc h^{\rm sd}_\lambda(\mc z,v,\tau)]},~~\text{for fixed}~u\ge0.   
\end{align}
Minimizing the rescaled action given in Eq.~\eqref{eq:PP_BMFT_action_scaled2}, one gets the saddle point solutions $\mc f^{\rm sd}_\lambda$ and $\mc h^{\rm sd}_\lambda$ that satisfy the following equations 
\begin{subequations}\label{eq:PP_saddle_BMFT}
\begin{align}
    &\partial_{\tau}\mc f^{\rm sd}_\lambda(\mc z,v,{\tau})+ v\partial_{\mc z}\mc f^{\rm sd}_\lambda(\mc z,v,{\tau})=0 \label{eq:sad_f_t}\\
    &\partial_{\tau}\mc h^{\rm sd}_\lambda(\mc z,v,{\tau})+ v\partial_{\mc z}\mc h^{\rm sd}_\lambda(\mc z,v,{\tau})=0\label{eq:sad_h_t}\\
    &\mc h^{\rm sd}_\lambda(\mc z,v,\msc T)=-\lambda \Theta(\mc z-u\msc T) \label{eq:sad_h_T}\\
    &\mc h^{\rm sd}_\lambda(\mc z,v,0)=-\lambda \Theta(\mc z)+\ln\Big(\frac{\mc f^{\rm sd}_\lambda(\mc z,v,0)}{\mc f_{\rm leq}(\mc z,v)}\Big),\label{eq:sad_h_0}
\end{align}
along with the normalization constant 
\begin{align}
    \int d\mc z \int dv~ \mc f^{\rm sd}_\lambda(\mc z,v,0)=N.
\end{align}
\end{subequations}
To obtain Eq.~\eqref{eq:sad_h_0}, we have taken the functional derivative of $\Delta \msc{F}^{\rm p}[\mc f(\mc z,v,0)]$ given in Eq.~\eqref{eq:PP_freeenergy_F_f_scaled}.
Solving Eq.~\eqref{eq:PP_saddle_BMFT} and evaluating $\msc{S}_{\lambda}[\mc f,\mc h]$ on these solutions, one finally gets the cumulant generating function 
\begin{align}
    \Gamma^{\rm p}_u(\lambda,\msc T) = - \lim_{\ell \to \infty}\frac{1}{\ell \msc T} \ln \big\langle e^{-\lambda \mc Q_u^{\rm p}(\ell\msc T)}\big\rangle= \frac{1}{\msc T}\msc{S}_{\lambda}[\mc f^{\rm sd}_\lambda(\mc z,v,\tau),\mc h^{\rm sd}_\lambda(\mc z,v,\tau)], \label{def:Gamma^p}
\end{align}
for fixed $u \ge 0$. Expanding this function in powers of $\lambda$ one can obtain various cumulants of $\mc Q_u^{\rm p}(\msf T)$.
To find the explicit expressions of these cumulants, one needs to solve the saddle point equations \eqref{eq:PP_saddle_BMFT}, which can be done easily. One finds the following solutions of the saddle point equations
\begin{subequations}
\label{sol:sad-eq-pp}
\begin{align}\label{eq:fxvt_evolve}
    \mc f^{\rm sd}_\lambda(\mc z,v,\tau)=\mc f^{\rm sd}_\lambda(\mc z-v\tau,v,0),~~{\rm and}~~\mc h^{\rm sd}_\lambda(\mc z,v,\tau)=\mc h^{\rm sd}_\lambda(\mc z-v\tau,v,0),
\end{align}
where 
\begin{align}
   \mc h^{\rm sd}_\lambda(\mc z,v,0)=-\lambda\Theta(\mc z+v\msc T-u\msc T).
\end{align}
and
\begin{align}\label{eq:fxv0_saddle}
    \mc f^{\rm sd}_\lambda(\mc z,v,0)=\mc f_{\rm leq}(\mc z,v)~{\rm e}^{-\lambda\big[\Theta(\mc z+v\msc T-u\msc T)-\Theta(\mc z)\big]}.
\end{align}
\end{subequations}
The generating function of  $\big\langle {\rm e}^{-\lambda \mc Q^{\rm p}_{-u}(\msf T)}\big\rangle$ can be evaluated by carrying out a similar saddle-point analysis. Following the same variational procedure as outlined above, the saddle-point solutions for the phase-space density and the auxiliary field, for this case, are again given by Eq.~\eqref{eq:fxvt_evolve} but now with
\begin{align}
\label{sol:sad-eq-pp_<}
    \mc h^{\rm sd}_\lambda(\mc z,v,0)=-\lambda\Theta(-z-v\msc T-u\msc T),~{\rm and}~\mc f^{\rm sd}_\lambda(\mc z,v,0)=\mc f_{\rm leq}(\mc z,v)~{\rm e}^{-\lambda\big[\Theta(-z-v\msc T-u\msc T)-\Theta(-z)\big]}.
\end{align}

\subsection{Distribution and cumulants for point particles}
Substituting the saddle point solutions from Eq.~\eqref{sol:sad-eq-pp}  to the action in Eq.~\eqref{eq:PP_BMFT_action_scaled2} and simplifying, we finally get the cumulant generating function $\Gamma^{\rm p}_u(\lambda,\msc T)$ defined in Eq.~\eqref{def:Gamma^p}. Hence, for large $\ell$, we have  
\begin{subequations}
\label{eq:CGF_BMFT_saddle0}  
\begin{align}
 \big\langle e^{-\lambda \mc Q_u^{\rm p}(\msf T)}\big\rangle \asymp e^{-\ell\msc T ~\Gamma^{\rm p}_u(\lambda,\msc T)}, \label{eq:CGF_BMFT_saddle1}  
\end{align}
where 
\begin{align}
\Gamma^{\rm p}_u(\lambda,\msc T) =\Big[\mf q_+(u,\msc T)\left(1-e^{-\lambda}\right)+\mf q_-(u,\msc T)\left(1-e^{\lambda}\right)\Big],
\label{eq:PP_scaledT_action_saddle}
\end{align}
with
\begin{align}\label{eq:PP_scaledT_action_saddle2}
\mf q_+(u,\msc T)= \int_{u}^\infty dv\int_{(u-v) }^{0}d\mc y~\mc f_{{\rm leq}}(\mc y\msc T,v),~~\mf q_-(u,\msc T)=\int_{-\infty}^u dv \int_{0}^{(u-v) }d\mc y~\mc f_{{\rm leq}}(\mc y\msc T,v).
\end{align}
\end{subequations}
Similarly, for the current in the negative direction, one finds
\begin{subequations}
\label{eq:CGF_BMFT_saddle0_<}  
\begin{align}
 \big\langle e^{-\lambda \mc Q_{-u}^{\rm p}(\msf T)}\big\rangle \asymp e^{-\ell \msc T ~\Gamma^{\rm p}_{-u}(\lambda,\msc T)}, \label{eq:CGF_BMFT_saddle1_<} 
\end{align}
for large  $\ell$, where 
\begin{align}
\Gamma^{\rm p}_{-u}(\lambda,\msc T) =   \Big[\mf q_-(-u,\msc T)\left(1-e^{-\lambda}\right)+\mf q_+(-u,\msc T)\left(1-e^{\lambda}\right)\Big].
\label{eq:PP_scaledT_action_saddle_<}
\end{align}
\end{subequations}

\noindent
Taking derivatives of the cumulant generating functions $\Gamma^{\rm p}_{\pm u}(\lambda,\msc T)$, one finds the following expression of the  cumulants 
\begin{align}\label{eq:Mean_Var_PP}
      \kappa_n^{\pm u}=\langle \mc Q_{\pm u}^{\rm p}(\msf T)^n \rangle_c=\msf T\big(\mf q_\pm(\pm u,\msc T)+(-1)^n \mf q_\mp(\pm u,\msc T)\big),~~\text{for}~n=1,2,...
\end{align}
The exponential form of the generating function in Eq.~\eqref{eq:CGF_BMFT_saddle0} implies that the distribution of $\mc Q^{\rm p}_{\pm u}(\msf T)$ satisfies a large deviation form 
\begin{align}\label{eq:PP_LDF_form}
    \mc P_{\rm p}(\mc Q^{\rm p}_{\pm u},\msf T)\asymp \exp\left(-  \msf T ~\msc H^{\rm p}_{u,\pm}\left(\frac{\mc Q^{\rm p}_{\pm u}}{\msf T},\frac{\msf T}{\ell}\right) \right),~~\text{for fixed}~u\ge 0,
\end{align}
where the rate function $\msc H^{\rm p}_{\pm u}(q,\msc T)$ can be calculated using the Legendre transform and one has
\begin{align}\label{eq:PP_Hq1}
   \msc H^{\rm p}_{u,\pm}(q,\msc T)=\left( \Gamma^{\rm p}_{\pm u}(\lambda,\msc T)-\lambda q\right)_{\lambda=\lambda^{*}_{\pm u}(q,\msc T)}
\end{align}
where $\lambda^{*}_{\pm u}(q,\msc T)$ is obtained by solving 
\begin{align}
    \dfrac{d}{d\lambda}\Gamma^{\rm p}_{\pm u}(\lambda,\msc T)\Big\vert_{\lambda=\lambda^{*}_{\pm u}}=q.
\end{align}
Following Eq.~\eqref{eq:PP_scaledT_action_saddle}, one gets
\begin{align} \label{eq:lambda*u}
\lambda^{*}_{\pm u}(q,\msc T)=-\ln\pa{\frac{q +\sqrt{q^2+4\mf q_\pm( \pm u,\msc T)\mf q_\mp(\pm u,\msc T)}}{2\mf q_\pm(\pm u,\msc T)}}.
\end{align}
Inserting the above form of $\lambda^*_\pm$ in Eq.~\eqref{eq:PP_Hq1} and simplifying, one gets
\begin{align}\label{eq:PP_Hq}
    \msc H^{\rm p}_{u,\pm}(q,\msc T)=\mf q_\pm(\pm u,\msc T)\Big[1-e^{-\lambda^*_{\pm u}(q,\msc T)}\big(\lambda^*_{\pm u}(q,\msc T)+1\big)\Big]+\mf q_\mp(\pm u,\msc T)\Big[1+e^{\lambda^*_{\pm u}(q,\msc T)}\big(\lambda^*_{\pm u}(q,\msc T)-1\big)\Big],
\end{align}
\noindent
where $\mf q_{\pm}(u,\msc T)$ are defined in Eq.~\eqref{eq:PP_scaledT_action_saddle2}.  For a homogeneous equilibrium state characterized by average phase space density $\mc f_{\rm leq}(\mc z,v)=\phi_0 \mc g(v)$, the quantities $\mf q_{\pm}(u,\msc T)$  in Eq.~\eqref{eq:PP_scaledT_action_saddle2} become independent of   $\msc T$ {\it i.e.} $\mf q_{\pm}(u,\msc T) \to q_\pm(u)$ where,
\begin{align}\label{eq:PP_scaledT_qpm_homogeneous}
   q_+(u)=\phi_0 \int_{u}^\infty dv~(v-u)~\mc g(v),~~q_-(u)=\phi_0\int_{-\infty}^u dv~ (u-v)~\mc g(v).
\end{align}
Consequently, $\lambda^{*}_{\pm u}(q,\msc T)$ in Eq.~\eqref{eq:lambda*u} and the large deviation form in Eq.~\eqref{eq:PP_LDF_form} become independent of $\msc T$. Hence, for homogeneous equilibrium state, the distribution  $\mc P_{\rm p}(\mc Q^{\rm p}_{\pm u},\msf T)$ satisfies the following large deviation form
\begin{align}\label{eq:PP_LDF_form}
    \mc P_{\rm p,eq}(\mc Q^{\rm p}_{\pm u},\msf T)\asymp \exp\left(-  \msf T ~\msc H^{\rm p,eq}_{u,\pm}\left(\frac{\mc Q^{\rm p}_{\pm u}}{\msf T}\right) \right),~~\text{for fixed}~u\ge 0.
\end{align}

\begin{figure}[htbp]
\centering
\includegraphics[width=7.77cm, height=6.5cm]{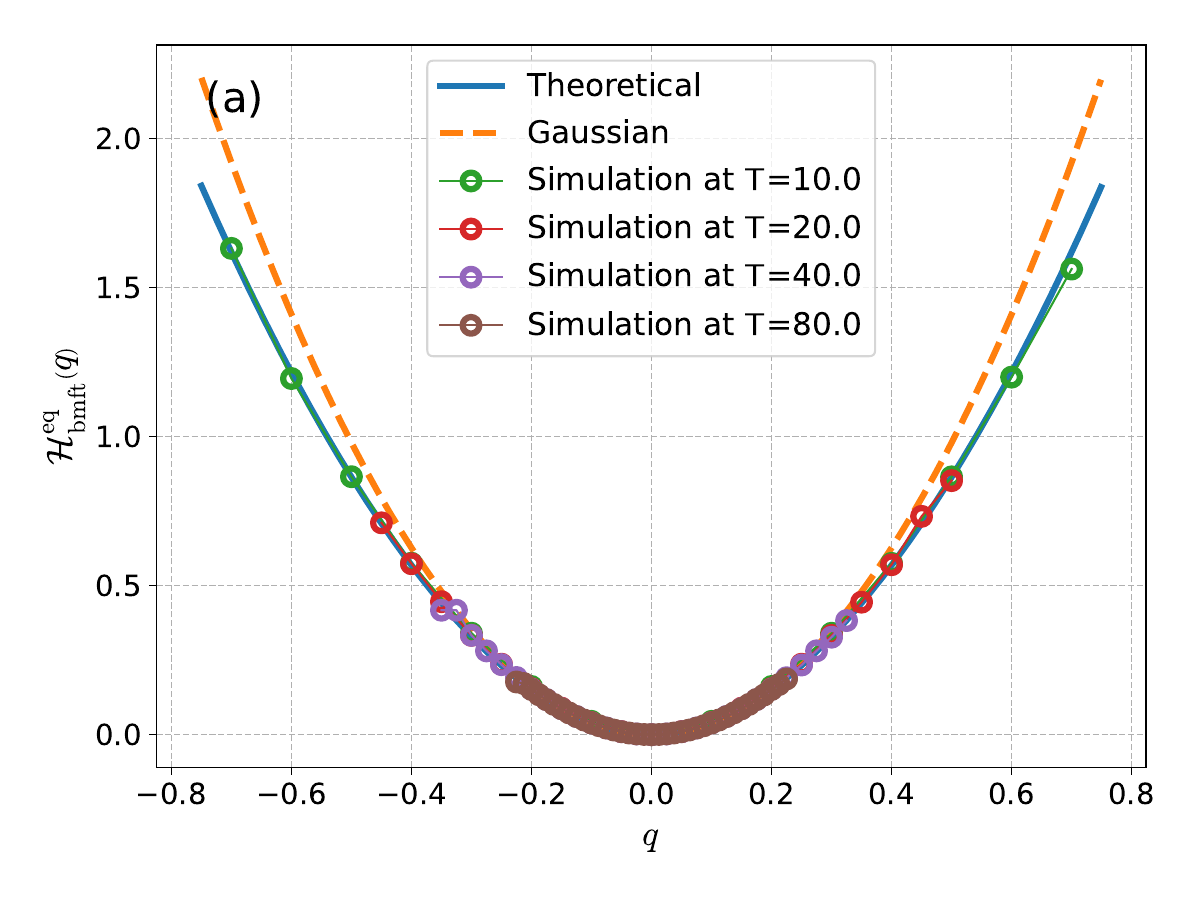}
\includegraphics[width=7.77cm, height=6.5cm]{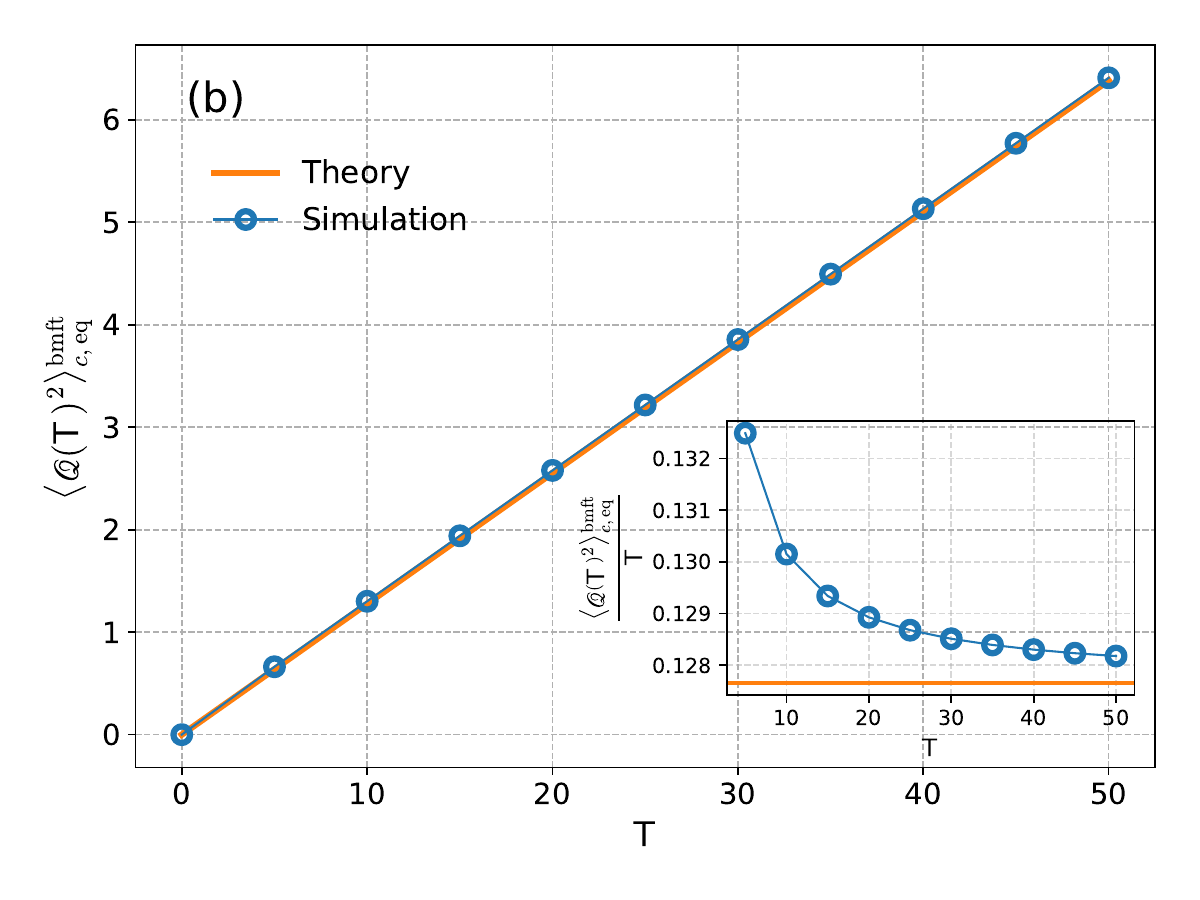}
\includegraphics[width=7.77cm, height=6.5cm]{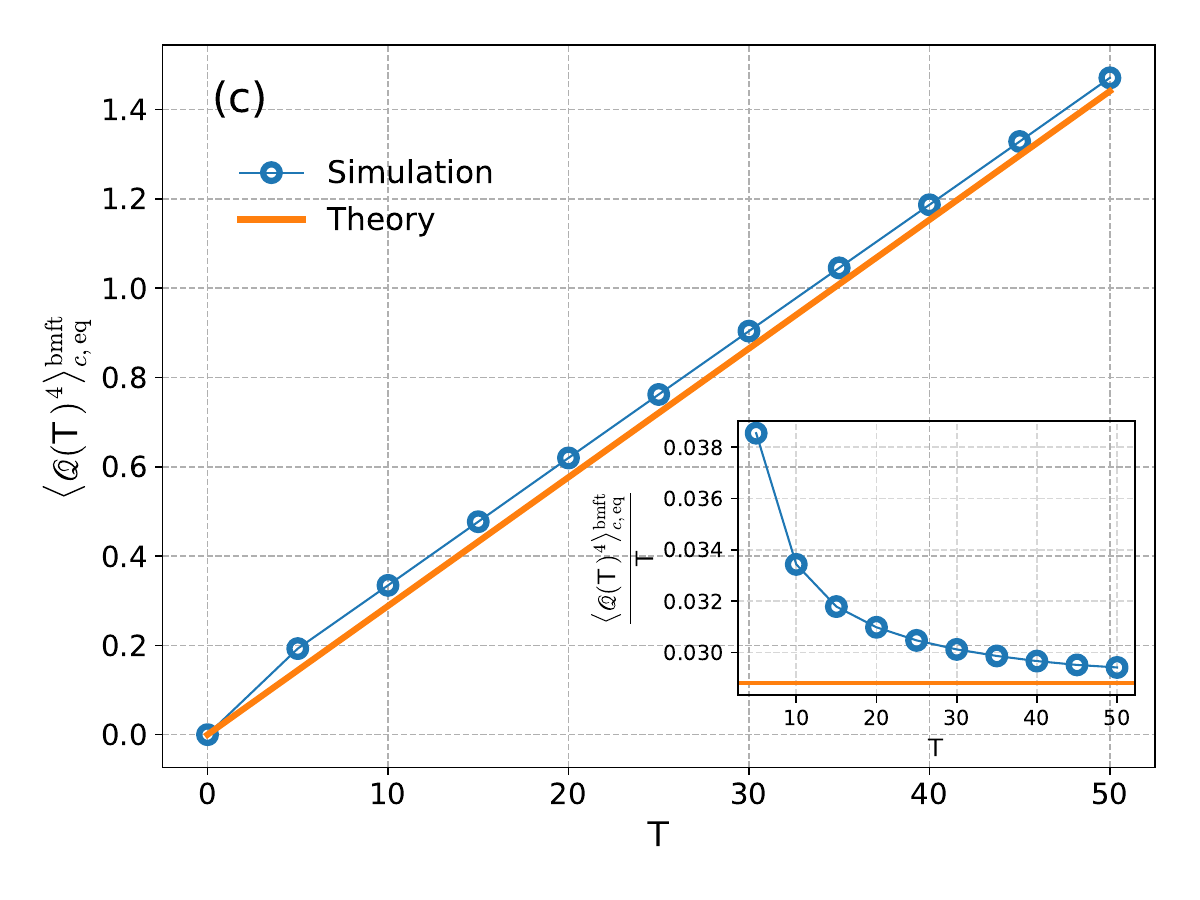}
\includegraphics[width=7.77cm, height=6.5cm]{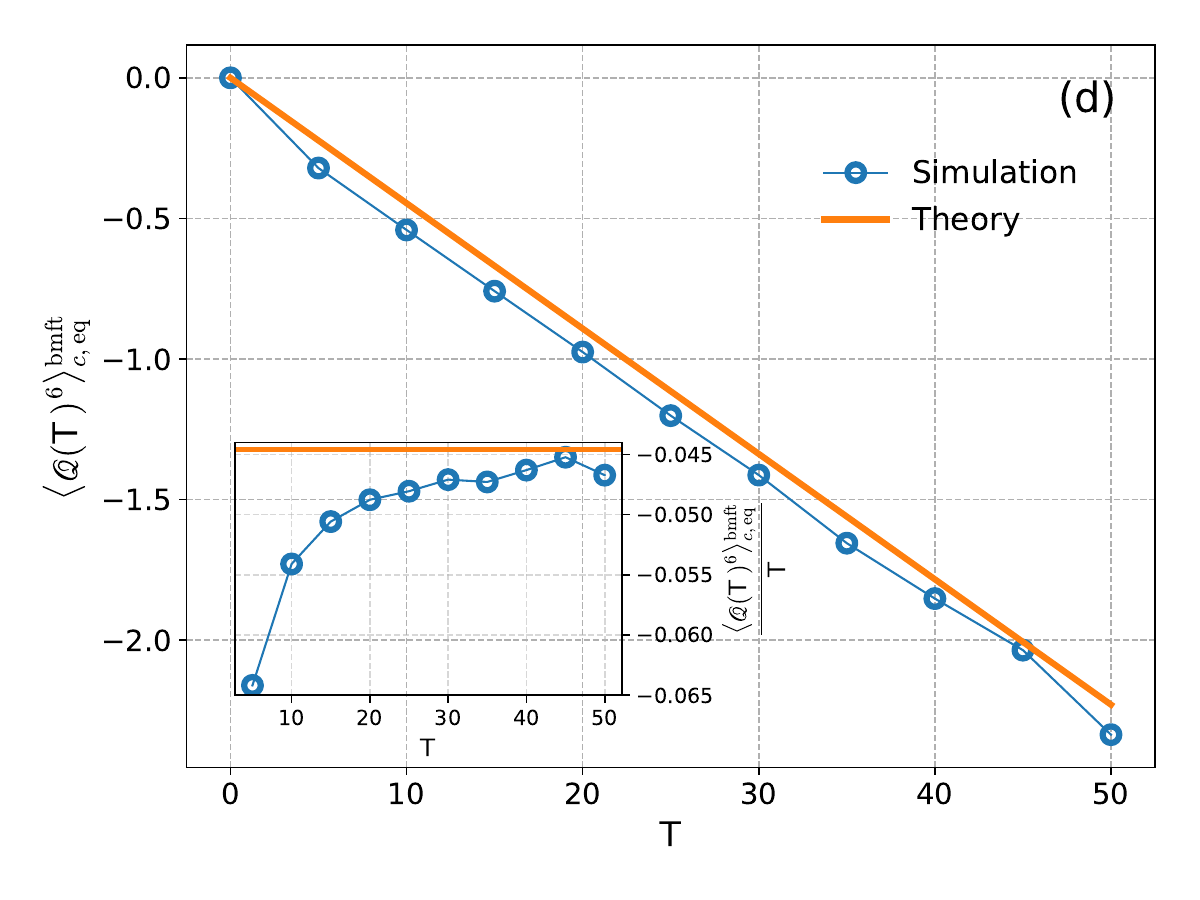}
\vspace*{-.5 cm}
\caption{(a) Comparison of the rate function $\msc H_{\rm bmft}(q)$ given in  Eq.~\eqref{eq: LDF_HR} (blue line) with the numerical simulation of the microscopic dynamics (circles) for an equilibrium gas of hard rods at times $\msf T=10,20,40,$ and $80$. The Gaussian rate function (orange dashed line) is constructed using the variance given in Eq.~\eqref{<Q^2>-BMFT}. While the typical fluctuations are well described by the Gaussian form, the atypical fluctuations in the tails exhibit clear deviations from Gaussian behavior. For the numerical simulation, the initial configurations are generated in the point-particle representation. For each $\msf T$, we consider $N=5 \msf{T}$ particles  uniformly distributed over the interval $[-L/2,L/2]$, with $L=20\msf{T}$. The resulting point-particle configurations are subsequently mapped onto the corresponding hard-rod configurations using Eq.~\eqref{eq:map_pp_rod}. The velocity distribution $\mc g(v)$ is taken to be a zero-mean Maxwellian with temperature $T_0=1$, and the rod length considered here is $a=1$.
(b) - (d) Comparisons of the second, fourth, and sixth cumulants, $\langle \mc Q(\msf T)^2 \rangle_c^{\rm bmft}, \langle \mc Q(\msf T)^4 \rangle_c^{\rm bmft}$, $\langle \mc Q(\msf T)^6 \rangle_c^{\rm bmft}$ respectively, obtained within BMFT using Eqs. \eqref{<Q^2>-BMFT}-\eqref{<Q^6>-BMFT} (orange line) with the numerical simulations (blue circles). The numerical simulations are performed with $L=20\msf{T}_{\rm max}$ with $\msf{T}_{\rm max}=50$ here, and $N=5\msf{T}_{\rm max}$, keeping all other parameters the same as in panel (a). For all cases, the numerical results are averaged over $5\times 10^9$ initial configurations.  Insets in (b)-(d) show the convergence to the asymptotic values for the cumulants. 
} 
\label{Fig:BMFT_LDF_cumulants}
\end{figure}

\subsection{Distribution and cumulants for hard rods}
In the previous section, we have shown that the distribution $\mc P_{\rm p}(\mc Q^{\rm p}_{\pm u},\msf T)$ of point particle current $\mc Q^{\rm p}_{\pm u}(\msf T)$ satisfies the large-deviation form as in Eq.~\eqref{eq:PP_LDF_form}, with the associated rate function $\msc H^{\rm p}_{u,\pm}\left(q,\msc T\right)$ given in Eq.~\eqref{eq:PP_Hq}. Given this distribution, the equivalence relations between the hard-rod and point-particle systems in Eqs.~\eqref{eq:PP_Rod_equivalences}-\eqref{eq:PP_Rod_equivalences_j_-ve}, together with Eq.~\eqref{eq:PP_Rod_dist_relation} allow us to obtain the distribution of the integrated current $\mc P_{\rm r}(\mc Q,\msf T)$ in the hard-rod system directly from its point-particle counterpart. Since $\mc P_{\rm p}(\mc Q^{\rm p}_{\pm u},\msf T)$ obeys the large-deviation form established above, it follows that the hard-rod current distribution, $\mc P_{\rm r}(\mc Q,\msf T)$, obtained from it through the mapping, likewise admits a large deviation form, 
\begin{align}\label{eq:Rod_LDF_form}
   \mc P_{\rm r}(\mc Q,\msf T) \asymp \exp\left(- \msf T ~\msc H_{\rm bmft}\left(\frac{\mc Q}{\msf T},\frac{\msf T}{\ell}\right) \right),
\end{align}
with  the rate function $\msc H_{\rm bmft}\left(\frac{\mc Q}{\msf T},\msc T\right)$ 
\begin{align}\label{eq: LDF_HR}
        \msc H_{\rm bmft}(q,\msc T) = \msc H^{\rm p}_{qa,+}(q,\msc T) \theta(q) + \msc H^{\rm p}_{|q|a,-}(-q,\msc T)  \theta(-q),
    \end{align}
where $\msc H^{\rm p}_{u,\pm}(q,\msc T)$ is given in Eq.~\eqref{eq:PP_Hq}.
Following Eqs.~\eqref{eq:lambda*u}-\eqref{eq:PP_Hq}, one can show that 
\begin{align}
    \lim_{q\to0^+}\msc H^{\rm p}_{qa,+}(q,\msc T)=\lim_{q\to0^-}\msc H^{\rm p}_{|q|a,-}(-q,\msc T),~
    {\rm and}~ \lim_{q\to0^+}\dfrac{d}{dq}\msc H^{\rm p}_{qa,+}(q,\msc T)=\lim_{q\to0^-}\dfrac{d}{dq}\msc H^{\rm p}_{|q|a,-}(-q,\msc T),
\end{align}
so that the rate function $\msc H_{\rm bmft}(q,\msc T)$ is continuous and indeed smooth across $q=0$. The results obtained so far are valid for an arbitrary initial profile. 

To compare with the statistics of the integrated current obtained using the CBFHD approach, which applies to a globally homogeneous gas of hard rods, we now specialize to the equilibrium gas of hard rods characterised by average phase space density $\mf f_{\rm eq} = \varrho_0 \mc g(v)$, where $\varrho_0$ represents the mass density and $\mc g(v)$ is the distribution from which the initial velocities of the rods are chosen independently. For this global equilibrium initial state, $\msc H_{\rm bmft}(q,\msc T)$ becomes independent of $\msc T$, {\it i.e.}, $\msc H_{\rm bmft}(q,\msc T) \to \msc H_{\rm bmft}^{\rm eq}(q)$, and one can show that
\begin{align}
\msc H_{\rm bmft}^{\rm eq}(q) = \msc H_{\rm bmft}^{\rm eq}(-q).
\end{align}
The asymptotic behaviour of the rate function $\msc H_{\rm bmft}^{\rm eq}(q)$ for small and large $|q|$  can be obtained from Eqs.~\eqref{eq: LDF_HR}-\eqref{eq:PP_Hq} and they strongly depend on the tail of the velocity distribution $\mc g(v)$. Assuming  Gaussian behavior at the tails  $g(v) \asymp e^{-v^2/2T_0}$, one finds
\begin{align}\label{eq:H_Small_large_q}
\msc H_{\rm bmft}^{\rm eq}(q) \approx
\begin{cases}
    \sqrt{\frac{\pi}{2}}~\frac{(1+a\phi_0)^2}{2\phi_0\sqrt{T_0}}q^2~~~~~~~\text{for~}q\to 0\\
    \frac{a^2}{2T_0}|q|^3~~~~~~~~~~~~~~\text{for~}~q\to\pm\infty.
\end{cases}
\end{align} 
A similar asymptotic form of the rate function $\sim q^3$, in the tails has also been reported earlier in the study of current fluctuations in the one-dimensional symmetric simple exclusion process (SSEP) with a step initial density profile \cite{Derrida2009}, in number fluctuations within the full eigenvalue statistics of the complex Ginibre ensemble \cite{Num_fluc_Kundu_2019}, and in the full counting statistics of two-dimensional rotating fermions \cite{Priyangshu_Fermions}. To obtain the cumulants of the hard-rod integrated current $\mc Q(\msf T)$, we consider the cumulant generating function 
\begin{align}
 \Gamma^{\rm r}_{\rm B}(\lambda)=-\lim_{\msf T\to \infty}\frac{1}{\msf T}\ln\big \langle {\rm e}^{-\lambda \mc Q(\msf T)}\big\rangle, 
 ~~~\text{where,}~~~    \big \langle {\rm e}^{-\lambda \mc Q(\msf T)}\big\rangle\asymp\int dq~{\rm e}^{-\msf T\big(\lambda q+\msc H_{\rm bmft}^{\rm eq}(q)\big)}.
\end{align}
For large $\msf T$, this integral is dominated by the saddle point $q^*(\lambda)$, that satisfies 
\begin{align}
    \frac{d}{dq}\msc H_{\rm bmft}^{\rm eq}(q)\big|_{q=q^*}=-\lambda.
\end{align}
The cumulant generating function is therefore obtained by evaluating the exponent at the saddle point $q=q^*(\lambda)$, giving
\begin{align}
    \Gamma^{\rm r}_{\rm B}(\lambda)=\big(\lambda q+\msc H_{\rm bmft}^{\rm eq}(q)\big)_{q=q^*}.
\end{align}
The current cumulants of the hard-rod system are subsequently computed from derivatives of $\Gamma^{\rm r}_{\rm B}(\lambda)$. For the equilibrium hard-rod gas of  density $\varrho_0 = \frac{\varrho_0}{1-a\varrho_0}$, and $\mc g(v)$ being a zero-mean Maxwellian velocity distribution  at temperature $T_0$, we find that all odd cumulants vanish, and the first three even cumulants are given explicitly by
\begin{align}
\langle \mc Q(\msf T)^2 \rangle_{c,\rm eq}^{\rm bmft}&=\msf T\sqrt{\frac{2}{\pi}}\frac{\phi_0\sqrt{T_0}}{(1+a\phi_0)^2}, \label{<Q^2>-BMFT}\\
\langle \mc Q(\msf T)^4 \rangle_{c,\rm eq}^{\rm bmft}&=\msf T\pa{\phi_0\sqrt{T_0}\sqrt{\frac{2}{\pi}}\frac{\pi-2a\pi\phi_0-3a^2(-4+\pi)\phi_0^2}{\pi\pa{1+a\phi_0}^6}}, \label{<Q^4>-BMFT}\\
\langle \mc Q(\msf T)^6 \rangle_{c,\rm eq}^{\rm bmft}&= \msf T \left[ \frac{\phi_0\sqrt{T_0} \sqrt{2}   \left(1020 a^4 \phi_0^4-150 \pi  a^2 \phi_0^2 (a\phi_0 +1) (3 a \phi_0 -1)\right)}{\pi ^{5/2} (a \phi_0 +1)^{10}}\right. \notag \\
&~~~~~~~~+ \left. \frac{\phi_0\sqrt{T_0} \sqrt{2} (\pi  a \phi_0 +\pi )^2 (3 a \phi_0  (15 a \phi_0 -8)+1)}{\pi ^{5/2} (a \phi_0 +1)^{10}} \right],\label{<Q^6>-BMFT}
\end{align}
where $\phi_0 = \frac{\varrho_0}{1-a\varrho_0}$.

The analytical expression of the large-deviation  function for the hard-rod gas, given in Eq.~\eqref{eq: LDF_HR}, is compared with the results of numerical simulations of the microscopic dynamics in Fig.~\ref{Fig:BMFT_LDF_cumulants}(a). The simulations are performed for an initial configuration of rods prepared in the point particle representation, with the distribution specified in Eq.~\eqref{eq:prob_init}. Specifically, $N=5\msf{T}$ particles are sampled uniformly from the interval $[-L/2,L/2]$, with $L=20\msf{T}$. The resulting point-particle configurations are subsequently mapped onto the corresponding hard-rod configurations using Eq.~\eqref{eq:map_pp_rod}. The velocity distribution is taken to be a zero-mean Maxwell distribution at temperature $T_0$. We find excellent agreement between the analytical prediction and the numerical data over the range considered. To assess the deviation of the current distribution from Gaussianity, we also compare the analytical rate function with the corresponding Gaussian rate function constructed using the variance given in Eq.~\eqref{<Q^2>-BMFT}.  Furthermore, the analytical expressions for the second, fourth, and sixth-order cumulants, given in Eqs.~\eqref{<Q^2>-BMFT}-\eqref{<Q^6>-BMFT}, are compared with the numerical results in Fig.~\ref{Fig:BMFT_LDF_cumulants} (b) - (d). The numerical simulations are performed with $L=20\msf{T}_{\rm max}$ with $\msf{T}_{\rm max}=50$ here, and $N=5\msf{T}_{\rm max}$, keeping all other parameters the same as in panel Fig.~\ref{Fig:BMFT_LDF_cumulants} (a). The analytical predictions show good agreement with the numerical simulations for all three cumulants, with insets showing the convergence to the asymptotic values. 
\begin{figure}[t]
\centering
\includegraphics[width=9.0cm, height=7.2cm]{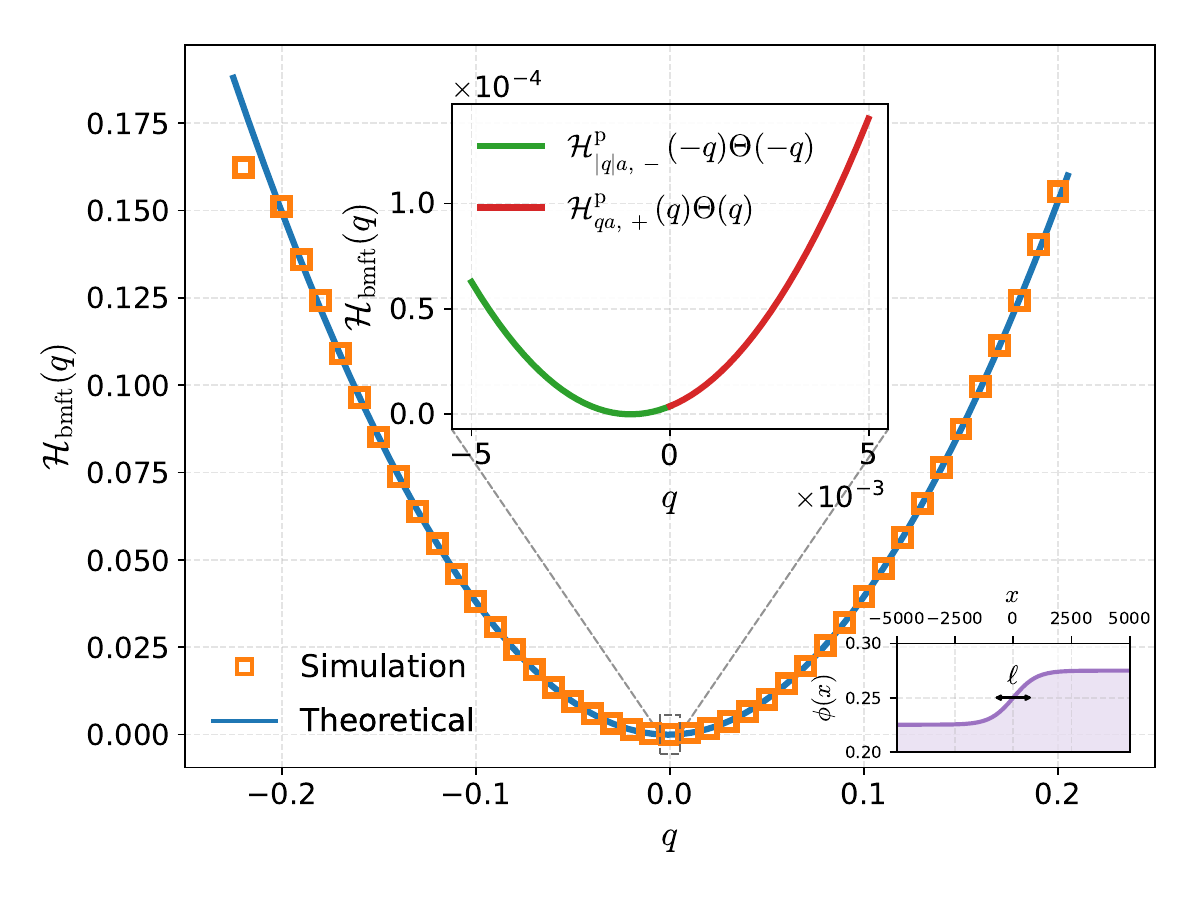}
\vspace*{-.2 cm}
\caption{ Plot comparing rate function $\msc H_{\rm bmft}(q)$ given in  Eq.~\eqref{eq: LDF_HR} (blue line) with the numerical simulation of the microscopic dynamics (circles) for an inhomogeneous initial profile at time $\msf T=100$. For the numerical simulation, the initial configurations are generated in point-particle representation. In particular, $N=2500$ point particles distributed on the interval $[-L/2,L/2]$. Their positions are sampled from the distribution $p(x)=\frac{4}{L}\Big(0.225+\frac{0.05}{1+\exp(-x/\ell)}\Big)$, with $L=10000$ and $\ell =500$. These positions are  subsequently mapped onto the corresponding hard-rod configurations using Eq.~\eqref{eq:map_pp_rod}. The velocity distribution is taken to be a zero-mean Maxwellian with temperature $T_0=1$. The rod length $a=1$. To obtain $H_{\rm bmft}(q)$ following  Eq.~\eqref{eq: LDF_HR} we have separately computed $\msc H^{\rm p}_{qa,+}(q) \theta(q)$ and $\msc H^{\rm p}_{|q|a,-}(-q)  \theta(-q)$ using Eq.~\eqref{eq:PP_Hq} and with the same form of $\phi(x)=Np(x)$. For this initial state, the distribution $\mc P_{\rm r}(\mc Q,T)$ and the corresponding  rate function $\msc H_{\rm bmft}(q)$ are expected to be asymmetric about the mean value where the mean should occur at negative $q$. This implies the minima of the large deviation function should occur at $q\ne0$, which is not clear from the main plot. To demonstrate that, we plot a zoomed version of the large deviation function  $\msc H_{\rm bmft}(q)$ near $q=0$ in the inset. The bottom-right inset shows the initial density profile. The numerical results are averaged over $10^8$ initial configurations. } 
\label{Fig:LDF_BMFT_Inhom}
\end{figure}
While Fig.~\ref{Fig:BMFT_LDF_cumulants}(a) presents the numerical verification of the theoretical predictions of the integrated current for an equilibrium gas of hard rods,  the expression for the rate function $\msc H_{\rm bmft}(q)$ in Eq.~\eqref{eq: LDF_HR} is, in fact, valid for an arbitrary initial profile. To demonstrate this numerically, we compare the analytical expression with  simulation data from the microscopic dynamics for an inhomogeneous initial condition in Fig.~\ref{Fig:LDF_BMFT_Inhom}. 
 We find excellent agreement between the numerical simulation and the analytical prediction.

\section{Integrated-current statistics using CBFHD}\label{CBFHD}
In this section, we employ the fluctuating hydrodynamic theory discussed in \cite{Chahal_tracer_2026} to determine the statistics of the integrated current $\mc Q(\msf T)$, defined in Eq.~\eqref{def:Q(T)}, for the equilibrium hard-rod gas characterized by mean phase space density $\mf f_{\rm eq}(X,v) = \varrho_0\mc g (v)$ where  $\varrho_0$ is the mass density and $\mc g(v)$ is the velocity  distribution As discussed in Sec.~\ref{sec:2}, the macroscopic
fluctuation of the phase space density $\mf f(X,v,t)$ for hard rods evolves according to the following stochastic partial differential equation \eqref{eq:FOFHD_eq1} \cite{Chahal_tracer_2026}.
As done in Sec.~\ref{sec:BMFT}, to characterize the statistics of $\mc Q(\msf T)$ we compute the generating function $\langle {\rm e}^{-\lambda \mc Q(\msf T)}\rangle$ which can be written as 
\begin{align}\label{eq:HR_MGF_FO1}
    \langle {\rm e}^{-\lambda \mc Q(\msf T)}\rangle=\int \mathcal{D}[\mf f(X,v,t)]~\msc{P}[\mf f(X,v,t)]~{\rm e}^{-\lambda \mc Q(\msf T)}.
\end{align}
where the functional $\msc{P}[\mf f(X,v,t)]$ represents the probability of the time evolution history of the profile $\mf f(X,v,t)$ over duration $0$ to $\msf T$.
Using the fluctuating hydrodynamic equation~\eqref{eq:FOFHD_eq1}, one can write 
\begin{align}\label{eq:HR_Pfxvt}
\msc{P}[\{\mf f(x,v,t);0\leq t \leq \msf T\}]\asymp \int d\mu ~{\rm e}^{\mu N}~&{\rm e}^{-\Delta\mathcal{F}^{\rm r}[\mf f(X,v,0)]} \delta \Big[\partial_t \mf f(X,v,t)+\msf V^{\rm eff}(v)\partial_X \mf f(X,v,t), \\
&-\frac{\mc D(v)}{2}\partial_X^2 \mf f(X,v,t) -\partial_X \mf f(X,v,t)\sqrt{\mc D(v)} \xi(v,t)\Big], \notag
\end{align}
where   the Lagrange multiplier $\mu$ ensures the normalization $\int dX\int dv~ \mf f(X,v,0)=N$. Here, $\Delta \mathcal{F}^{\rm r}[\mf f]=\mathcal{F}^{\rm r}[\mf f]-\mathcal{F}^{\rm r}[\mf f_{\rm eq}]$, where the free energy functional for the hard rods \cite{Kundu_2025}
\begin{align}\label{eq:HR_freeenergy_F_f}
    \mathcal{F}^{\rm r}[\mf f(0)]=\int dX\int dv~\mf f(X,v,0)~\Bigg[\omega(v)+\mu+\ln\Big(\frac{\mf f(X,v,0)}{1-a\varrho(X,0)}\Big)\Bigg],
\end{align}
with $\omega(v)=-\ln\varrho_0-\ln\mc g(v)$ and $\varrho(X,0)=\int dv~\mf f(X,v,0)$. The equilibrium density  $\mf f_{\rm eq}(X,v)=\varrho_0 \mc g(v)$ would minimize the free energy functional $\mathcal{F}^{\rm r}[\mf f(0)]$. Substituting $\msc{P}[\mf f(X,v,t)]$ from Eq.~\eqref{eq:HR_Pfxvt} into Eq.~\eqref{eq:HR_MGF_FO1}, $\langle {\rm e}^{-\lambda \mc Q(\msf T)}\rangle$ becomes
\begin{align}\label{eq:HR_MGF_FO2}
    \langle {\rm e}^{-\lambda \mc Q(\msf T)}\rangle=\int \mathcal{D}[\mf f(X,v,t)]&~{\rm e}^{-\Delta \mathcal{F}^{\rm r}[\mf f(0)]} ~{\rm e}^{-\lambda \mc Q(\msf T)}
    \Big\langle \delta \Big[\partial_t \mf f(X,v,t)+\msf V^{\rm eff}(v)\partial_X \mf f(X,v,t)\cr
&~~~~-\frac{\mc D(v)}{2}\partial_X^2 \mf f(X,v,t) -\partial_X \mf f(X,v,t)\sqrt{\mc D(v)} \xi(v,t)\Big]\Big\rangle_{\xi}
\end{align}
where the angular bracket with subscript $\xi$ denotes the averaging over the history of noise within the window $[0,\msf T]$. Using the integral representation of the delta functional through an auxiliary field $\mf h(x, v, t)$ and computing the average with respect to the noise $\xi$ in Eq.~\eqref{eq:HR_MGF_FO2}, one finds 
\begin{align}\label{eq:FOFHD_MGF}
   \langle {\rm e}^{-\lambda \mc Q(\msf T)}\rangle=\int d\mu ~{\rm e}^{\mu N}\int \mathcal{D}[\mf f(X,v,t),\mf h(X,v,t)]~{\rm e}^{-\mathcal{S}_{\lambda}[\mf f(X,v,t), \mf h(X,v,t)]}
\end{align}
where the action $\mathcal{S}_{\lambda}[\mf f(X,v,t), \mf h(X,v,t)]$ is given by
\begin{align}\label{eq:HR_FOFHD_action}
    \mathcal{S}_{\lambda}[\mf f, \mf h]&=\Delta \mathcal{F}^{\rm r}[\mf f(0)]+\lambda\int dX\int dv~\big[\mf f(X,v,\msf T)-\mf f(X,v,0)\big]\Theta(X) \\
    &~~+\int_0^\msf Tdt \int dX\int dv ~\mf h(X,v,t)\Big[\partial_t\mf f(X,v,t)+\msf V^{\rm eff}(v)\partial_X\mf f(X,v,t)-\frac{\mc D(v)}{2}\partial^2_X\mf f(X,v,t)\Big]\cr
    &~~-\frac{1}{2}\int_0^\msf Tdt\int dv\int du\int dX\int dY~\partial_X\mf h(X,v,t)\mf f(X,v,t)~\partial_{Y}\mf h(Y,u,t)\mf f(Y,u,t)\mathcal{G}(u,v),\notag
\end{align}
where recall the noise covariance $\mc{G}(v,u)$ from Eq.~\eqref{eq:G_uv}. 
We assume that, in the long-time limit, the action in Eq.~ \eqref{eq:HR_FOFHD_action} scales linearly with time $\msf T$, which will be justified a posteriori. Consequently, the generating function can be evaluated in the large $\msf T$ limit using a saddle point approximation.
Minimizing the action in Eq.~\eqref{eq:HR_FOFHD_action} one gets the saddle point solutions $\mf f^{\rm sd}_\lambda$ and $\mf h^{\rm sd}_\lambda$ that satisfy the following equations
\begin{subequations}\label{eq:saddle_olla}
\begin{align}
  & \partial_t \mf f^{\rm sd}_\lambda(X,v,t)+\msf V^{\rm eff}(v)\partial_X\mf f^{\rm sd}_\lambda(X,v,t)-\frac{\mc D(v)}{2}\partial^2_X\mf f^{\rm sd}_\lambda(X,v,t)=-\big(\partial_X\mf f^{\rm sd}_\lambda(X,v,t)\big)~\msc W^{\rm sd}_\lambda(v,t)\label{eq:saddle_olla1}\\
  & \partial_t\mf h^{\rm sd}_\lambda(X,v,t)+\msf V^{\rm eff}(v)\partial_X\mf h^{\rm sd}_\lambda(X,v,t)+\frac{\mc D(v)}{2}\partial^2_X\mf h^{\rm sd}_\lambda(X,v,t)=-\partial_X\big(\mf h^{\rm sd}_\lambda(X,v,t)\big)\msc W^{\rm sd}_\lambda(v,t)\label{eq:saddle_olla2}\\
  &\mf h^{\rm sd}_\lambda(X,v,\msf T)=-\lambda\Theta(X)\label{eq:saddle_olla3}\\
  &\mf h^{\rm sd}_\lambda(X,v,0)=-\lambda\Theta(X)+\Big(\frac{\delta(\Delta \mathcal{F}^{\rm r}[\mf f(0)])}{\delta \mf f(X,v,0)}\Big)_{\mf f(0)=\mf f^{\rm sd}_\lambda(0)}\label{eq:saddle_olla4}\\
  &{\rm where}~~\msc W^{\rm sd}_\lambda(v,t)=\int dY\int du~\mf f^{\rm sd}_\lambda(Y,u,t)\mathcal{G}(u,v)\partial_Y\mf h^{\rm sd}_\lambda(Y,u,t)\label{eq:saddle_olla5}.
\end{align}
along with the normalization condition
\begin{align}
    \int dX\int dv~ \mf f^{\rm sd}_\lambda(X,v,0)=N.
\end{align}
\end{subequations}
The functional derivative appearing in Eq.~\eqref{eq:saddle_olla4} can be evaluated from the hard-rod free-energy functional in Eq. \eqref{eq:HR_freeenergy_F_f}. In terms of the equilibrium phase space density $\mf f_{\rm eq}(X,v)$ that minimizes the free energy functional $\mathcal{F}^{\rm r}[\mf f(0)]$ subject to the normalization constraint, we can write
\begin{align}\label{eq:e^(-DelF)}
    \Delta \mathcal{F}^{\rm r}[\mf f(0)]&=\int dX\int dv\Bigg[ \mf f(X,v,0)\ln\Big(\frac{\mf f(X,v,0)}{\mf f_{\rm eq}(X,v)}\Big)-  \mf f(X,v,0)\ln\Big(\frac{1-a\varrho(X,0)}{1-a\varrho_{\rm eq}(X)}\Big)\cr
    &~~~~~~~~~~~~~~~~~~~~~-\frac{1}{1-a\varrho_{\rm eq}(X)}\big(\mf f(X,v,0)-\mf f_{\rm eq}(X,v)\big)\Bigg],
\end{align}
where $\varrho(X,0)=\int dv\mf f(X,v,0)$, $\varrho_{\rm eq}(X)=\int dv\mf f_{\rm eq}(X,v)$. Thus,
\begin{align}\label{eq:FOFHD_var_DelF_f}
   \frac{\delta(\Delta \mathcal{F}^{\rm r}[\mf f(0)])}{\delta \mf f(X,v,0)}= \ln\Bigg(\frac{\mf f(X,v,0)}{\mf f_{\rm eq}(X,v)}\Bigg)-\ln\Bigg(\frac{1-a\varrho(X,0)}{1-a\varrho_{\rm eq}(X)}\Bigg)+\frac{1}{1-a\varrho(X,0)}-\frac{1}{1-a\varrho_{\rm eq}(X)}
\end{align}
The saddle point action following Eq.~\eqref{eq:saddle_olla} is given by
\begin{align}\label{eq:S_saddle_Olla}
    \mathcal{S}_{\lambda}[\mf f^{\rm sd}_\lambda,\mf h^{\rm sd}_\lambda]=&\Delta\mathcal{F}^{\rm r}[\mf f^{\rm sd}_\lambda(0)]+\lambda \mc Q(\msf T)|_{\mf f^{\rm sd}_\lambda,\mf h^{\rm sd}_\lambda}\cr 
    &~~~~+\frac{1}{2}\int_0^{\msf T}dt\int dX\int dv ~\mf f^{\rm sd}_\lambda(X,v,t)~\partial_X \mf h^{\rm sd}_\lambda(X,v,t) \msc W^{\rm sd}_\lambda(v,t)
\end{align}
where $\Delta \mathcal{F}^{\rm r}[\mf f]=\mathcal{F}^{\rm r}[\mf f]-\mathcal{F}^{\rm r}[\mf f_{\rm eq}]$ and $\mc Q(\msf T)|_{\mf f^{\rm sd}_\lambda,\mf h^{\rm sd}_\lambda}$ is the current $\mc Q(\msf T)$ evaluated at $\mf f^{\rm sd}_\lambda(X,v,t)$ and $\mf h^{\rm sd}_\lambda(X,v,t)$  in Eq.~\eqref{def:Q(T)}. The cumulant generating function $\msc G_{\rm F}(\lambda,\msf T) = - \ln \langle {\rm e}^{-\lambda \mc Q(\msf T)}\rangle$ 
is then obtained by substituting the saddle point solutions $\left(\mf f^{\rm sd}_\lambda(X,v,t),\mf h^{\rm sd}_\lambda(X,v,t)\right)$  in Eq.~\eqref{eq:FOFHD_MGF} and we get
\begin{align}\label{eq:FO_CGF_Define}
 \msc G_{\rm F}(\lambda,\msf T)\asymp \mathcal{S}_{\lambda}[\mf f^{\rm sd}_\lambda,\mf h^{\rm sd}_\lambda].
\end{align}
where $\mathcal{S}_{\lambda}[\mf f^{\rm sd}_\lambda,\mf h^{\rm sd}_\lambda]$ is given in Eq.~\eqref{eq:S_saddle_Olla}. Solutions of the saddle point equations are presented in Appendix~\ref{app:sol-saddle-olla}. Evaluating the action $\mathcal{S}_{\lambda}[\mf f^{\rm sd}_\lambda,\mf h^{\rm sd}_\lambda]$ explicitly and taking the large $\msf T$ limit, as shown  in Appendix~\ref{app:action-large-T-FO} in detail, we find that the cumulant generating function $\msc G_{\rm F}(\lambda,\msf T)$is proportional to $\msf T$  to  leading order of $\msf T$ {\it i.e.,}
\begin{align}\label{eq:FO_Gamma_scaleT-mt}
   \msc G_{\rm F}(\lambda,\msf T)=\msf T \Gamma_{\rm cb}^{\rm r}(\lambda).
\end{align}
with
\begin{align}\label{eq:FO_gamma_lambda-mt}
    \Gamma_{\rm cb}^{\rm r}(\lambda)=&\int dZ\Big[\frac{\rho_{\rm eq}(Z)}{1-a\rho_{\rm eq}(Z)}-\frac{\rho^{\rm sd}_\lambda(Z,0)}{1-a\rho^{\rm sd}_\lambda(Z,0)}\Big]\cr
    &~~~~~~~+\frac{\lambda^2}{2}\int dv\int du~F(-\msc V^{\rm eff}_\lambda(v),v)F(-\msc V^{\rm eff}_\lambda(u),u)\mc G(u,v).
\end{align}
where $F(-\msc V^{\rm eff}_\lambda(v),v)$ is given in Eq.~\eqref{eq:FO_larget_F_function2} and $\rho^{\rm sd}_\lambda(Z,0)$ is given in Eq.~\eqref{eq:FO_rhoX0_scaled}. To evaluate these quantities, one needs to evaluate the function $\msc V^{\rm eff}_\lambda(v)$ (defined in Eq.~\eqref{eq:FO_new eff_vel}) for a given $\lambda$ which can be obtained by solving  equation~\eqref{eq:FO_larget_W_sd} self consistently. 

\subsection{Distribution of the integrated current  and cumulants}
The generating function of the integrated current $\mc Q(\msf T)$ following Eqs.~\eqref{eq:FO_CGF_Define} and \eqref{eq:FO_Gamma_scaleT-mt}, takes the form
\begin{align}
    \langle {\rm e}^{-\lambda \mc Q(\msf T)}\rangle={\rm e}^{-\msf T\Gamma_{\rm cb}^{\rm r}(\lambda)},
\end{align}
in the large time limit where $\Gamma_{\rm cb}^{\rm r}(\lambda)$ is given in Eq.~\eqref{eq:FO_gamma_lambda-mt}. Consequently, the probability distribution of $\mc Q(\msf T)$ obeys the large-deviation form,
\begin{align}\label{eq:FO_LDF_form}
    \mc P_{\rm r}(\mc Q,\msf T)\asymp \exp\left(- \msf T ~\msc H_{\rm cb}\left(\frac{\mc Q}{\msf T}\right) \right).
\end{align}
where the corresponding rate function can be obtained using the Legendre duality between the cumulant generating function $\Gamma_{\rm cb}^{\rm r}(\lambda)$ and $\msc H_{\rm cb}(\frac{\mc Q}{\msf T})$, and can be given as
\begin{align}\label{eq:FO_HQ1}
   \msc H_{\rm cb}\left(q\right)=\Bigg(\lambda q-\Gamma_{\rm cb}^{\rm r}(\lambda)\Bigg)_{\lambda=\lambda^{*}(q)}
\end{align}
with $\lambda^{*}(q)$ is determined from
\begin{align}\label{eq:FO_rate_lambda*}
    \dfrac{d}{d\lambda}\Gamma_{\rm cb}^{\rm r}(\lambda)\bigg|_{\lambda=\lambda^{*}}=q,
\end{align}
for real $q$.
An explicit analytical evaluation of the rate function is in general difficult. We therefore determine it numerically via the iterative procedure described below. The expression for $\Gamma_{\rm cb}^{\rm r}(\lambda)$ involves the effective velocity $\msc V^{\rm eff}_\lambda(v)$ which is determined self-consistently through $\msf{W}^{\rm sd}_\lambda(v)$, as given by Eq. \eqref{eq:FO_larget_W_sd}, together with Eq.~\eqref{eq:FO_new eff_vel}. 
Since $F\big(-\msc V^{\rm eff}_\lambda(v),v\big)$ in Eq.~\eqref{eq:FO_larget_F_function} itself depends on the effective velocity through the saddle-point density $\rho^{\rm sd}_\lambda(Z)$, these equations form a self-consistent set. For each fixed value of $\lambda$, we solve this system iteratively as follows. Starting from an initial guess $\msf{W}^{\rm sd}_{0}(v)$, we first calculate the corresponding effective velocity $\msc V^{\rm eff}_{\lambda,0}(v)$. This determines the saddle point density $\rho^{\rm sd}_0(Z)$ following Eqs.~\eqref{eq:FO_saddle_scaled_H_lambda} and \eqref{eq:FO_rhoX0_scaled}. The resulting density is then used to construct $F_0\big(-\msc V^{\rm eff}_{\lambda,0}(v),v\big)$. The update rule for the next iterate is then
\begin{align}
    \msf{W}^{\rm sd}_{n+1}(v)=-\lambda\int du~F_n(-\msc V^{\rm eff}_{\lambda,n}(u),u)\mc G(u,v).
\end{align}
This procedure is repeated until the difference between two successive iterations satisfies the condition $|\msf{W}^{\rm sd}_{n+1}(v)-\msf{W}^{\rm sd}_{n}(v)|<\epsilon$, where $\epsilon$ is a prescribed numerical tolerance. The converged solution of $\msf{W}^{\rm sd}_\lambda(v)$ then simultaneously satisfies the self-consistency relation and determines the corresponding saddle-point density $\rho^{\rm sd}_\lambda(Z)$ and $F\big(-\msc V^{\rm eff}_\lambda(v),v\big)$. For the converged solution, the cumulant generating function $\Gamma_{\rm cb}^{\rm r}(\lambda)$ is then evaluated following Eq.~\eqref{eq:FO_gamma_lambda-mt}. Repeating this procedure for a range of $\lambda$ yields the full cumulant generating function, and the rate function $\msc H_{\rm cb}\left(q\right)$ is obtained from Eq.~\eqref{eq:FO_HQ1}. For the homogeneous equilibrium hard rod gas considered here, with density $\varrho_0$, and a symmetric velocity distribution $\mc g(v)$, the optimal $\lambda^{*}(q)$ for a given value of $\frac{\mc Q}{\msf T}=q$ is determined from Eq.~\eqref{eq:FO_rate_lambda*}, and the rate function follows from Eq.~\eqref{eq:FO_HQ1}. 

We next turn to the cumulants of the integrated current. 
The current cumulants are then obtained from its derivatives at $\lambda=0$. More explicitly, assuming $\Gamma_{\rm cb}^{\rm r}(\lambda)$ admits the expansion
\begin{align}
    \Gamma_{\rm cb}^{\rm r}(\lambda)=\lambda \msc J_1+\frac{\lambda^2}{2!}\msc J_2+\frac{\lambda^3}{3!}\msc J_3+\frac{\lambda^4}{4!}\msc J_4+\cdots,
\end{align}
where $\msc J_n=(-1)^n\frac{\partial^n\Gamma_{\rm cb}^{\rm r}(\lambda)}{\partial \lambda^n}\big|_{\lambda=0}$, the $n$-th cumulant is obtained from the $n$-th coefficient. For an equilibrium gas of hard rods considered in this section, with velocities sampled from a normalized symmetric distribution {\it i.e.} for $\mc g(v)= \mc g(-v)$, all odd cumulants vanish. We therefore first consider the second cumulant and subsequently discuss the fourth cumulant. An explicit expression for the second cumulant is obtained as (see Appendix \ref{sec:2nd_cum_FOFHD} for derivation)
\begin{align}\label{eq:FO_Q2_expression}
    \langle \mc Q(\msf T)^2 \rangle_{c,\rm eq}^{\rm cb}=\msf T\frac{\varrho_0\sqrt{T_0}}{\sqrt{\pi}(1-a\varrho_0)}\Big[\sqrt{2}-(2a-a^2\varrho_0)\varrho_0(\sqrt{2}-1)+ a^2\rho_0^2\Big].
\end{align}
\begin{figure}[h]
\centering
\includegraphics[width=9.0cm, height=7.2cm]{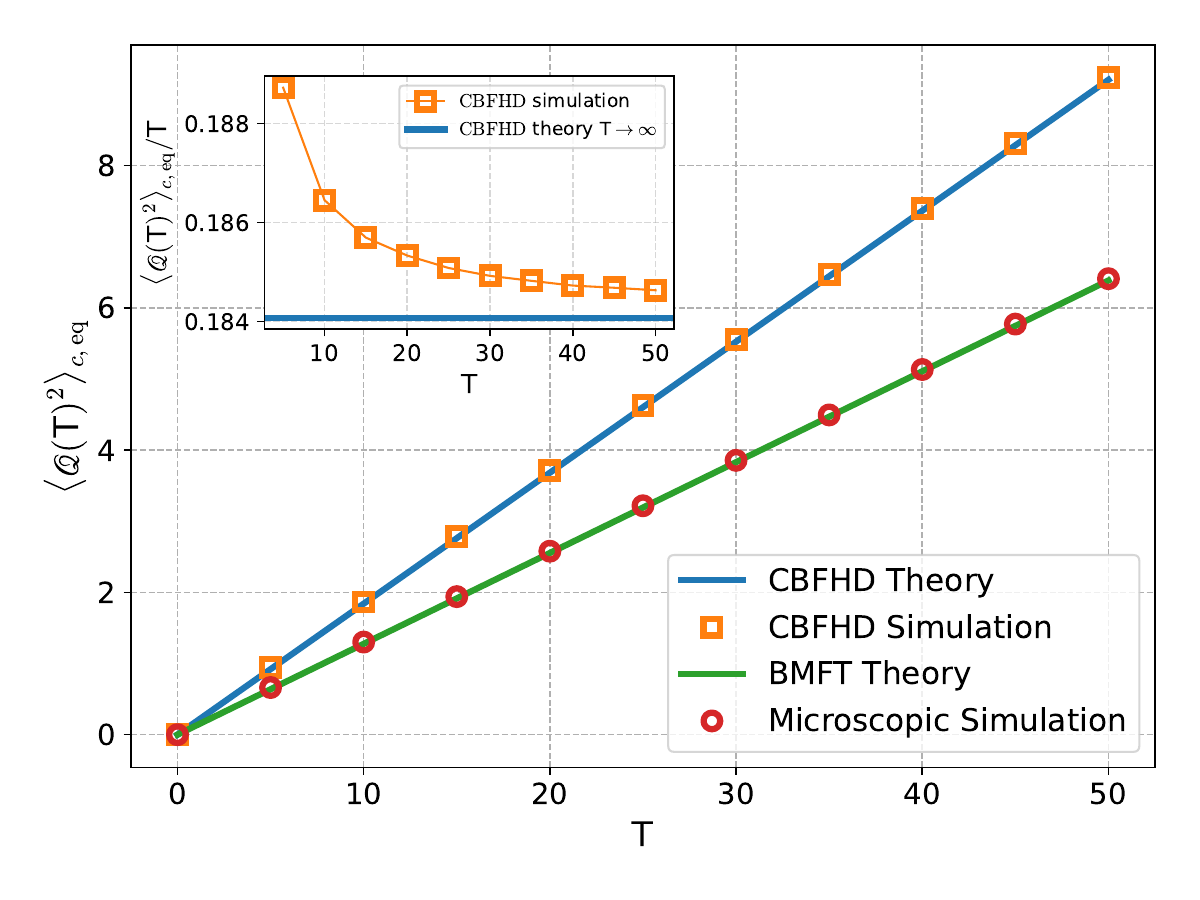}
\vspace*{-.2 cm}
\caption{Comparison of the second cumulant $\langle \mc Q(\msf T)^2\rangle_{c,\rm{eq}}^{\rm cb}$ obtained from the CBFHD approach, as in Eq.~\eqref{eq:FO_Q2_expression} (blue line), with numerical simulations (squares). The simulations are performed for a homogeneous gas of hard rods with density $\varrho_0=0.2,$ rod length $a=1$, and $N=1640$ rods. The velocities are sampled from a zero-mean Maxwellian distribution with temperature $T_0=1$. The rods evolve according to the effective Brownian dynamics in Eq.~\eqref{Brownian-QP}, with an effective drift, but are driven by correlated noises as given in Eq.~\eqref{eq:noise_corr}. The inset shows the convergence of the cumulant to its asymptotic value. For comparison, the second cumulant obtained from the BMFT approach, following Eq.~\eqref{<Q^2>-BMFT} (green line), is shown together with numerical simulations of the microscopic hard-rod dynamics (circles). The two theoretical predictions are shown to be distinct, $\langle \mc Q(\msf T)^2\rangle_{c,\rm{eq}}^{\rm bmft}\neq\langle \mc Q(\msf T)^2\rangle_{c,\rm{eq}}^{\rm cb}$, and the CBFHD result does not agree with the microscopic hard-rod simulations (red circles). The microscopic simulation has been performed for $a=1$, $N=1640$ and $\varrho_0=0.2$, following the similar procedure described in Fig.~\ref{Fig:BMFT_LDF_cumulants}. For both panels, the averages are taken over $10^9$ independent initial configurations.
} 
\label{Fig:FOFHD_variance}
\end{figure}
\begin{figure}[t]
\centering
\includegraphics[width=7.74cm, height=6.5cm]{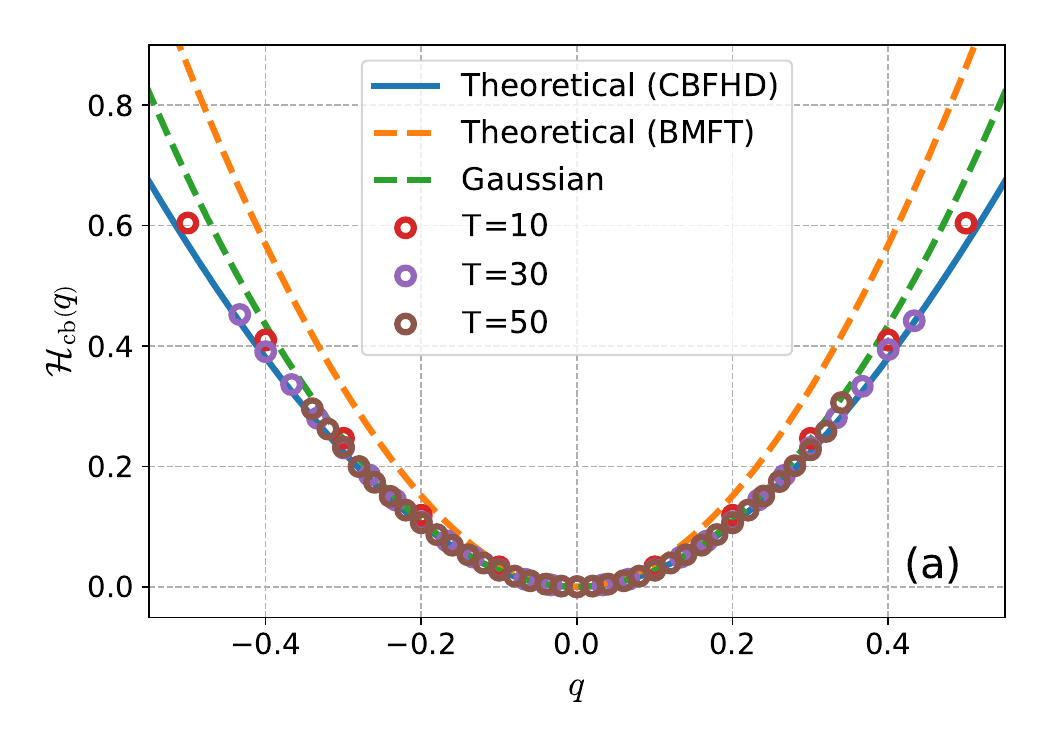}
 \includegraphics[width=7.74cm, height=6.5cm]{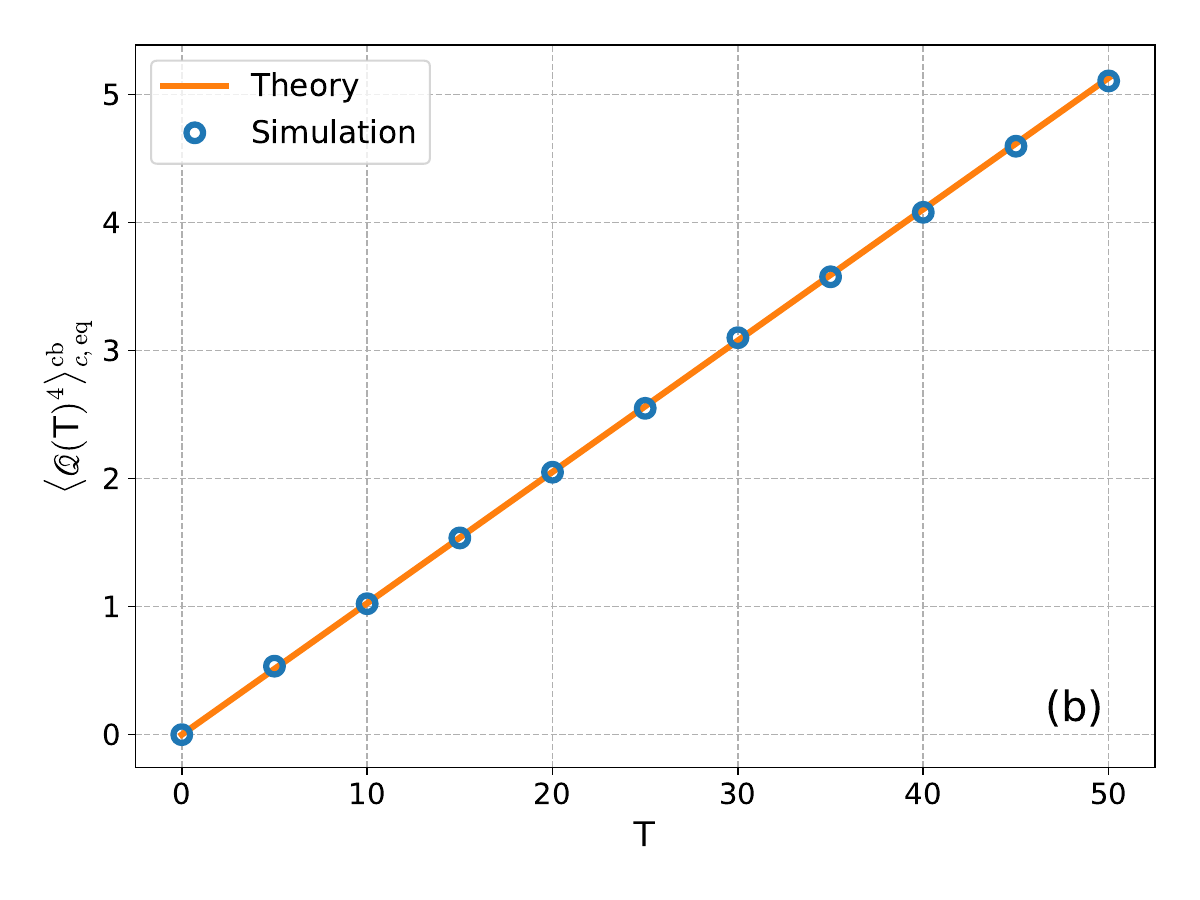}
\vspace*{-.6 cm}
\caption{(a) Comparison of the rate function $\msc H_{\rm cb}(q)$ (blue solid line), obtained by iteratively solving Eq.~\eqref{eq:FO_gamma_lambda-mt} with the numerical simulation for a homogeneous hard-rod gas of density $\varrho_0=0.2$, rod length $a=1$ at times $\msf T=10, 30,50$ (circles).The simulations are performed for $N=1650$, with velocities sampled from a zero-mean Maxwellian distribution at temperature $T_0=1$. The numerical results are obtained by evolving the hard rods according to the effective Brownian dynamics in Eq.~\eqref{Brownian-QP}, with an effective drift, but are driven by correlated noises as given in Eq.~\eqref{eq:noise_corr}. The Gaussian distribution,  constructed using Eq.~\eqref{eq:FO_Q2_expression}, is shown for comparison (green dashed line). While the typical fluctuation agrees with the Gaussian distribution, they exhibit distinct behavior in the tails. The rate function $\msc H_{\rm bmft}(q)$, given in  Eq.~\eqref{eq: LDF_HR}, is also plotted (orange dashed line), showing that the two theoretical predictions are distinct.  (b) Comparison of the fourth cumulant $\langle \mc Q(\msf T)^4\rangle_{c,{\rm eq}}^{\rm cb}$ obtained within the CBFHD framework by iteratively solving  Eq.~\eqref{eq:FO_gamma_lambda-mt} (solid line) 
with the numerical simulation (circles). The simulations use the same dynamics and parameters as in panel (a). For both panels, the numerical results are averaged over $10^9$ initial configurations. 
} 
\label{Fig:FOFHD_LDF}
\end{figure}
We find that the variance of the integrated current grows linearly with time $\msf T$. The exact expression for the second cumulant of the integrated current obtained within the CBFHD framework, in Eq.~\eqref{eq:FO_Q2_expression}, is compared with numerical simulations in Fig.~\ref {Fig:FOFHD_variance}. The simulations are performed for a homogeneous gas of hard rods with density $\varrho_0$ that effectively move as Brownian particles following the dynamics in Eq. \eqref{Brownian-QP} with an effective drift, but are driven by correlated noises as given in Eq.~\eqref{eq:noise_corr}. 
The numerical results are found to be in excellent agreement with the analytical expression in Eq. \eqref{eq:FO_Q2_expression}. The inset shows the convergence of the cumulant to its asymptotic value.  This agreement confirms the internal consistency of the CBFHD description for the stochastic hydrodynamic model. It is also worthy to compare the second cumulant obtained from the CBFHD approach with that obtained from the BMFT formulation given in Eq. \eqref{<Q^2>-BMFT}, and with direct numerical simulations of the microscopic hard-rod dynamics. We find that the two expressions are distinct, {\it i.e.},
\begin{align}
\langle \mc Q(\msf T)^2 \rangle_{c,{\rm eq}}^{\rm bmft} \neq \langle \mc Q(\msf T)^2 \rangle_{c,{\rm eq}}^{\rm cb},
\end{align}
and, crucially, the CBFHD prediction does not agree with the microscopic hard-rod simulations, whereas the BMFT prediction does. This discrepancy is clearly demonstrated in Fig.~\ref{Fig:FOFHD_variance}. In Fig.~\ref{Fig:FOFHD_LDF}(a), we compare the rate function obtained from the CBFHD framework with numerical simulations. The CBFHD rate function is obtained by solving the relation in Eq.~\eqref{eq:FO_gamma_lambda-mt} iteratively for a homogeneous hard-rod gas of density $\varrho_0$. The numerical results are obtained for the hard rods, evolving according to the effective Brownian dynamics in Eq.~\eqref{Brownian-QP} with correlated noises as specified in Eq.~\eqref{eq:noise_corr}. We find good agreement between the CBFHD prediction and the numerical data. The resulting rate function also exhibits clear non-Gaussian behavior. For comparison, the Gaussian rate function constructed using the variance predicted by Eq.~\eqref{eq:FO_Q2_expression} is shown in the same figure. Furthermore, to check amount of non-Gaussianity we compute the fourth cumulant theoretically and compare with numerical simulation in Fig.~\ref{Fig:FOFHD_LDF}(b).

To determine the fourth cumulant from the cumulant generating function in Eq.~\eqref{eq:FO_gamma_lambda-mt}, a direct analytical expansion of $\Gamma_{\rm cb}^{\rm r}(\lambda)$ to fourth order in $\lambda$ becomes cumbersome. We therefore evaluate $\Gamma_{\rm cb}^{\rm r}(\lambda)$ numerically by solving the self-consistency problem for each fixed value of $\lambda$, following the iterative procedure described above. We then evaluate $\Gamma_{\rm cb}^{\rm r}(\lambda)$ for a set of values of  $\lambda$ and extract the fourth-order coefficient, or equivalently the fourth derivative of $\Gamma_{\rm cb}^{\rm r}(\lambda)$ at $\lambda=0$. We compare the resulting fourth cumulant with numerical simulations performed for the homogeneous hard-rod gas evolving according to the effective Brownian dynamics Eq.~\eqref{Brownian-QP}. As shown in Fig.~\ref{Fig:FOFHD_LDF}(b), the fourth cumulant obtained from CBFHD shows good agreement with the numerical simulations.

\section{Conclusion} \label{sec:conclusion}
In this paper, we have studied the statistics of the integrated current in a one-dimensional equilibrium hard-rod gas using two recently proposed fluctuating hydrodynamic frameworks: (i) the Ballistic Macroscopic Fluctuation Theory (BMFT)~\cite{doyon2023ballistic, Kundu_2025} and (ii) the fluctuating hydrodynamic theory of correlated Brownian motion discussed in \cite{Chahal_tracer_2026}, whose predictions do not agree.  The theoretical predictions of both frameworks have been derived analytically and compared systematically against direct numerical simulations of the microscopic hard-rod dynamics. This comparison establishes a quantitative correspondence between the microscopic and macroscopic descriptions of current fluctuations, and reveals clear differences between the two hydrodynamic formulations. In particular, the integrated current provides a natural and sensitive probe for distinguishing between BMFT and CBFHD, and for assessing the ability of each framework to capture large-scale fluctuations in the hard-rod gas in the context of full counting statistics.

We have first studied the current statistics employing BMFT. By formulating the BMFT calculation in the hard-point-particle representation, we derived the full distribution of the integrated current for the hard-rod system and demonstrated explicitly that it takes a large-deviation form. An exact expression for the associated rate function was obtained analytically valid for a certain class of initial states characterized by arbitrary initial average phase-space density profile. In particular for global equilibrium state, we found  current cumulants up to sixth order explicitly and shown to grow linearly in time, in excellent agreement with direct numerical simulations of the microscopic hard-rod dynamics. The computation of the current distribution $\msc P(\mc Q(\msf T))$ has been carried out by mapping this problem to equivalent current statistics problem in point particle gas. This mapping would remain valid even if the rods move stochastically ({\it e.g.,} Brownian motion) instead of ballistic motion. In particular for Brownian hard rod gas one can straightforwardly use macroscopic fluctuation theory instead of BMFT as  has been applied recently \cite{jangid2026MFT}. 
In the point particle picture one would get the same Poisson structure for the cumulant generating function $\Gamma^{\rm p}_{u}(\lambda, \msc T)$ in Eq.~\eqref{eq:CGF_BMFT_saddle0}, except now the functions $\mc q_\pm(u,\msc T)$ get modified with diffusive scaling with respect to time. 

We subsequently analyzed the current statistics within the CBFHD framework and demonstrated that the current distribution likewise obeys a large-deviation principle. While an exact expression for the corresponding rate function is difficult to compute in our case, we determined it numerically via the iterative self-consistency procedure, as described in Sec.~\ref{CBFHD}. For the cumulants, an exact analytical expression for the second cumulant was derived within CBFHD, whereas the fourth cumulant was evaluated numerically. In both frameworks, the variance of the integrated current is found to grow linearly in time; however, the proportionality coefficients predicted by the two theories are distinct. 
Furthermore, we have shown that in both frameworks, the distribution of the integrated current admits a Large deviation form under ballistic scaling and theoretically computed the associated rate functions. Crucially, while the BMFT predictions for the variance and the rate function are in excellent agreement with direct numerical simulations of the microscopic hard-rod dynamics, the CBFHD prediction is not. Instead, the CBFHD result is consistent with numerical simulations of the effective correlated Brownian dynamics. 

Taken together, these results demonstrate that the integrated current serves as a sensitive observable for systematically distinguishing between the two hydrodynamic frameworks in the hard-rod system. Our findings suggest that CBFHD fails to provide an accurate description of large-scale current fluctuations in the hard-rod gas, whereas BMFT yields predictions that are in quantitative agreement with the microscopic dynamics, at least in the context of full counting statistics. At the same time, the agreement between CBFHD and the effective correlated Brownian dynamics confirms the consistency of the CBFHD description for the stochastic hydrodynamic model itself.

\section{Acknowledgements}
AK acknowledges the financial support under project ANRF/ARGM/2025/001207/MTR from the ANRF, DST, Government of India. TM, IM, and AK would also like to acknowledge the support from the DAE, Government of India, under Project No. RTI4001.
\appendix
\addtocontents{toc}{\fixappendix}

\section{Details of the initial distribution in Eq.~\eqref{eq:prob_init}}
\label{app:P_ini}
Given the number configuration $(N_1,N_2=N-N_1)$ on the two sides of the origin, the resulting joint probability density for the initial positions and velocities of the $N$ hard-point particles is therefore given by
 \begin{align}
 \begin{split}
 \mathfrak{P}_{N_1}(\{x_{j},v_{j}\},0)=N!&\prod_{j=-N_1}^{N_2-1} \Omega(x_{j},v_{j})\prod_{j=- N_1}^{-1}\Theta(x_{j+1}-x_j)\Theta(-x_1) \Theta(x_0) \\ 
 &\times~\prod_{j=0}^{N_2-2}\Theta(x_{j+1}-x_j) \Theta(x_{-N_1}+L_1) \Theta(L_2-x_{N_2-1}),
 \end{split}
 \label{eq:prob_init-a}
 \end{align}
where  $\Theta(x)$ is the Heaviside theta function and $N_1+N_2=N$. The product over Heaviside $\Theta$ functions ensures the ordering $\{x_{j+1}\ge x_{j}~;~j=-N_1,\dots, N_2-1\}$. Note that the joint probability in Eq.~\eqref{eq:prob_init-a} can be written as 
\begin{align}
   \mathfrak{P}_{N_1}(\{x_{j},v_{j}\},0)= \binom{N}{N_1} \mc q_{l}^{N_1} \mc q_r^{N-N_1}~\times~\mathfrak{P}_{N_1}^{l}(\{x_{j},v_{j}\},0) ~\times ~\mathfrak{P}_{N-N_1}^{r}(\{x_{j},v_{j}\},0),
\end{align}
where, recall, 
$\mc q_{l} = \int_{-L_1}^0dx \int dv~ \Omega(x,v)$ and $\mc q_r = \int_0^{L_2}dx \int dv ~\Omega(x,v)$. Here,   $\mathfrak{P}_{N_1}^{l}(\{x_{j},v_{j}\},0)$ represents the joint probability of the positions and velocities of $N_1$ particles in the left part $[-L_1,0]$ and is given by
\begin{align}
 \mathfrak{P}_{N_1}^{l}(\{x_{j},v_{j}\},0)=N_1!\prod_{j=-N_1}^{-1} \Omega_l(x_j,v_j)~\prod_{j=- N_1}^{-2}\Theta(x_{j+1}-x_j)\Theta(x_{-N_1}+L_1)\Theta(-x_{-1})
 ,\label{eq:prob_init_l}
\end{align}
where $\Omega_l(x,v) = \frac{\Omega(x,v)}{\mc q_l}$. Similarly,  the joint probability of the positions and velocities of $N_2$ particles on the right is given by 
\begin{align}
 \mathfrak{P}_{N_2}^{r}(\{x_{j},v_{j}\},0)=N_2!\prod_{j=0}^{N_2-1} \Omega_r(x_j,v_j)~\prod_{j=0}^{N_2-2}\Theta(x_{j+1}-x_j)\Theta(L_2-x_{N_2-1})\Theta(x_0)
 ,\label{eq:prob_init_r}
\end{align}
with $\Omega_r(x,v) = \frac{\Omega(x,v)}{\mc q_r}$.

\section{Integrated-current distribution: Microscopic approach}\label{sec:app_micro}
In this section, we calculate the distribution of the integrated current $\mc P_{\rm r}(\mc Q,\msf T)$ for the hard-rod gas using a microscopic approach. 
We first focus on the computation of $\mc P_{\rm r}(\mc Q,\msf T)$ for $\mc Q >0$; The computation for $\mc Q <0$ can be done following  an entirely analogous procedure. 

For $\mc Q >0$, recall that the equivalence relation reads
\begin{align}\label{eq:micro_PP_Rod_equivalences}
   \mathbb{P}^c_{\rm r}(\mc Q(\msf T)>j)=P^{\rm r}(X_{-j}(\msf T)>0)= P^{\rm p}(x_{-j}(\msf T)>(j-1/2)a), 
\end{align}
for $j\geq0$. Here $P^{\rm p}(x_{-j}(\msf T)>z)$ denotes the probability that the point particle $-j$ lies right to the position $z$ at time $\msf T$ and can be expressed as 
\begin{align}\label{eq:micro_cum_prob_S}
P^{\rm p}(x_{-j}(\msf T)>z)=\int_{z}^{\infty}dx~\mt P_{-j}(x,\msf T),
\end{align}
where $\mt P_{-j}(x,\msf T)$ denotes the probability density for finding the point particle labeled $-j$ at position $x$ at time $\msf T$.
The distribution $\mc P_{\rm r}(\mc Q=j,\msf T)$, for $j\geq0$ can then be obtained from this point-particle result using Eq.~\eqref{eq:PP_Rod_dist_relation}.

We first compute the probability of finding the point particle labeled $-j$ at position $x$ at time $t$. This probability can be written as 
\begin{align}
P^{\rm p}(x_{-j}(\msf T)>z)= \sum_{N_1=0}^N\msc P^{\rm p}_{\rm c}(x_{-j}(\msf T)>z|N_1, N-N_1) ~\times~ \left[ \binom{N}{N_1} \mc q_{l}^{N_1} \mc q_r^{N-N_1}\right], \label{def: P^p_micro}
\end{align}
where $\msc P^{\rm p}_{\rm c}(x_{-j}(\msf T)>z|N_1,N_2)$ is the probability that $x_j(\msf T) >z$ conditioned to start from a partition   $(N_1,N_2=N-N_1)$ and the last factor within square brackets represents the probability of such a partition.

To compute the probability $\msc P^{\rm p}_{\rm c}(x_{-j}(\msf T)>z|N_1,N_2)$, we consider two distinct possibilities: (i) the particle found at $x$ at time $t$ originated from among the $N_1$ particles initially located in the left half of the system, whose contribution we denote by $\mathcal L_{j}(x,N_1,N_2,t)$; and (ii) it originated from among the $N_2$ particles initially located in the right half, whose contribution we denote by $\mathcal R_{j}(x,N_1,N_2,t)$. Summing these two contributions gives the probability that the point particle labeled $-j$ is found at position $x$ at time $t$, for this particular initial configuration. To evaluate the contribution from process (i) we first consider the probability of finding a point particle at $x$ at time $t$, given that it was initially located on the left side of the origin. This contribution is given by
\begin{align}\label{eq:pl_expression}
p_l(x,t)
=\int_{-L_1}^{0}dy~G(x,t|y,0)\phi_l(y).
\end{align}
where $\phi_l(x)= \frac{\int dv~\Omega(x,v)}{\mc q_l}$ and  $G(x,t|y,0) = \int dv ~\delta(x-y -vt) \mc g(v) =\frac{1}{t} \mc g\left(\frac{x-y}{t} \right)$ represents the propagator of a point particle to reach $x$ at time $t$ starting from $y$. Similarly, for particles initially located on the right side of the origin, the corresponding contribution reads
\begin{align}\label{eq:pr_expression}
p_r(x,t)
=\int_{0}^{L_2}dy~G(x,t|y,0)\phi_r(y),
\end{align} 
where $\phi_r(x)= \frac{\int dv~\Omega(x,v)}{\mc q_r}$.
In terms of these probabilities in Eqs.~\eqref{eq:pl_expression}-\eqref{eq:pr_expression}, $\mathcal L_{j}(x,N_1,N_2,t)$ can be written as 
\begin{align}\label{eq:micro_Lj}
    \mathcal L_{j}(x,N_1,N_2,t)=\sum_{n_l+n_r=N_1-j}\binom{N_1-1}{n_l}&N_1 p_l(x,t)[q_l(x,t)]^{n_l}[1-q_l(x,t)]^{N_1-1-n_l}\cr
    &\times
    \binom{N_2}{n_r}[q_r(x,t)]^{n_r}[1-q_r(x,t)]^{N_2-n_r},
\end{align}
where
\begin{align}\label{eq:micro_qrl}
    q_{r,l}(x,t) = \int^{x}_{-\infty}d w~ p_{r,l}(w,t).
\end{align}
The contribution $\mathcal L_{j}(x,N_1,N_2,t)$ admit the following interpretation. Out of the $N_1$ particles initially located to the left of the origin, one particle can be placed at $x$ at time $t$ with probability $p_l(x,t)$ in $N_1$ possible ways. The factors $[q_l(x,t)]^{n_l}$ and $[1-q_l(x,t)]^{N_1-1-n_l}$ give the probability that, of the remaining $(N_1-1)$ particles initially on the left, $n_l$ end up to the left of $x$ and $(N_1-1-n_l)$ to its right at time $t$. Similarly the factors $[q_r(x,t)]^{n_r}$ and [1-$q_r(x,t)]^{N_2-n_r}$ represent the probability that, of the  $N_2$ particles initially on the right, $n_r$ end up to the left of $x$ and $N_2-n_r$ to its right. The constraint $n_l+n_r=(N_1-j)$ ensures that exactly $(N_1-j)$ particles lie to the left of $x$ at time $t$ -- the condition required for the particle at $x$ to be precisely the one labeled $-j$. The contribution from process (ii) is obtained by a similar argument, and $\mathcal R_{j}(x,N_1,N_2,t)$ can accordingly be expressed as
\begin{align}\label{eq:micro_Rj}
    \mathcal R_{j}(x,N_1,N_2,t)=\sum_{n_l+n_r=N_1-j}\binom{N_1}{n_l}&N_2 p_r(x,t)[q_l(x,t)]^{n_l}[1-q_l(x,t)]^{N_1-n_l}\cr
    &\times
    \binom{N_2-1}{n_r}[q_r(x,t)]^{n_r}[1-q_r(x,t)]^{N_2-1-n_r}.
\end{align}
\begin{figure}[t]
\centering
\includegraphics[width=9.0cm, height=7.2cm]{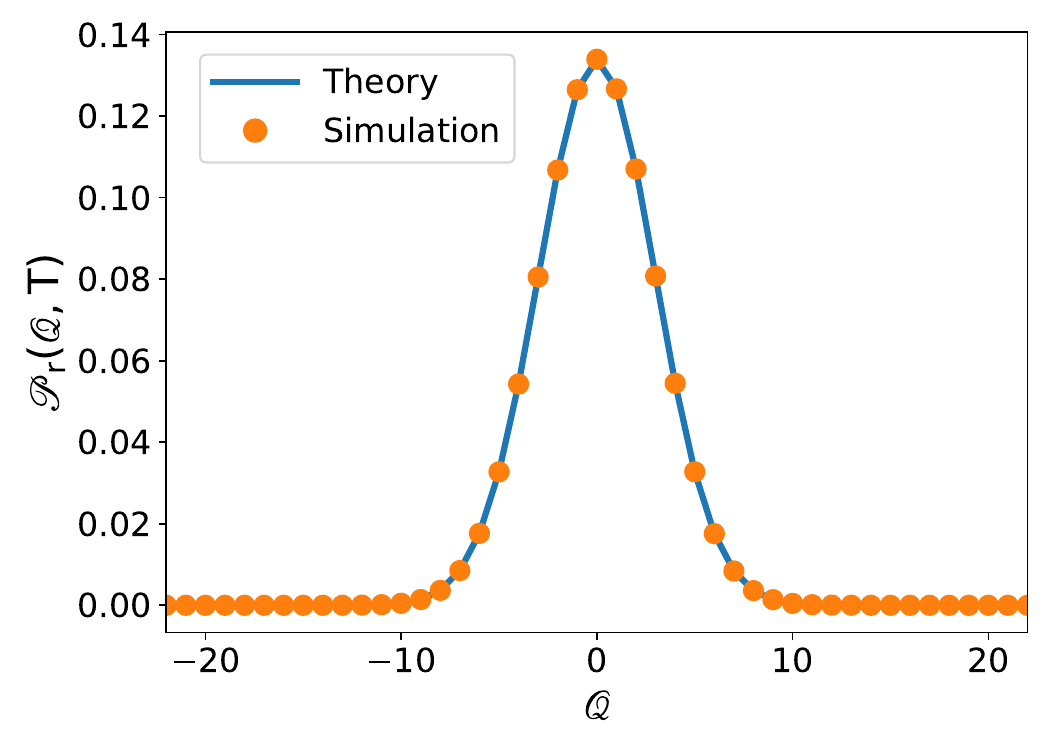}
\vspace*{-.2 cm}
\caption{ Plot comparing the probability distribution of integrated current $\mc P_{\rm r}(\mc Q,\msf T)$ between the analytical exact expression (blue line) following Eqs.~\eqref{eq:PP_Rod_dist_relation} and \eqref{eq:micro_Prob_Pp}  and the numerical
simulation (orange points) for an equilibrium gas of hard rods with density $\varrho_0=1/3$ at time $\msf T=50.$  For the numerical simulation, initial configurations are generated in the point-particle representation with point-particle density $\phi_0=0.5$.  $N=200$ particles sampled uniformly from the interval $[-L,L]$ with $L=200$ and subsequently mapped to the hard-rod configurations using Eq.~\eqref{eq:map_pp_rod}. The velocity distribution is taken to be a zero-mean Maxwellian at temperature $T_0=1$, and the rod length $a=1$. The numerical results are averaged over $10^7$ initial configurations.}
\label{Fig:Micro_dist}
\end{figure}
Therefore, for a general initial configuration we have 
\begin{align}\label{eq:micro_Prob_Pp}
   \msc  P^{\rm p}(x_{-j}(\msf T)>z|N_1,N_2)=\int_{z}^{\infty}dx~\big[\mathcal L_{j}(x,N_1,N_2,\msf T)+\mathcal R_{j}(x,N_1,N_2,\msf T)\big],
\end{align}
inserting which in Eq.~\eqref{def: P^p_micro} we get the probability $P^{\rm p}(x_{-j}(\msf T)>z)$.
The distribution $\mc P_{\rm r}(\mc Q,\msf T)$ of the integrated current for hard rods is then obtained  according to
Eq.~\eqref{eq:PP_Rod_dist_relation} for $\mc Q>0$, while the distribution for $\mc Q<0$ can be computed by an entirely analogous analysis following Eq.~\eqref{eq:PP_Rod_equivalences_j_-ve}. This result holds for an arbitrary initial mass-density profile $\phi(x)$. We compare the analytical prediction for $\mc P_{\rm r}(\mc Q,\msf T)$ with numerical simulations of an equilibrium hard-rod gas of density $\varrho_0$. For an equilibrium state, the corresponding point-particle density is spatially homogeneous, {\it i.e.}, $\phi(x)=\phi_0$, where $\phi_0$ is related to the hard-rod density through $\phi_0=\frac{\varrho_0}{1-a\varrho_0}$. For this homogeneous state, one obtains
\begin{align}
    \mc P_{\rm r}(\mc Q,\msf T)=\mc P_{\rm r}(-\mc Q,\msf T).
\end{align}
For the numerical simulations, we first generate the initial configurations in the point-particle representation, with $N=200$ particles sampled uniformly from the interval $[-L,L]$, $L=200$. These are subsequently mapped onto the corresponding hard-rod configurations using Eq.~\eqref{eq:map_pp_rod}. The particle velocities are sampled independently from a zero-mean Maxwellian distribution with temperature $T_0=1$. As shown in Fig.~\ref{Fig:Micro_dist}, we find good agreement between the analytical prediction and the numerical results. \\

\noindent
{\bf Remark:} The above microscopic calculations can straightforwardly be extended for Brownian hard rods with only change being the propagator $G(x,t|y,0)$ now has a diffusive scaling {\it i.e.,} $G(x,t|y,0) = \frac{1}{\sqrt{t}}\mc g\left(\frac{x-y}{\sqrt{t}} \right)$.

\section{Details of the derivation of $\mc P_{\rm r}(\mc Q,\msf T)$ using CBFHD}
\label{app:FOFHD-derv}
In this section we provide details of the computation of the saddle point calculation, large deviation function and the cumulant of the integrated current $\mc Q(\msf T)$ using the correlated Brownian fluctuating hydrodynamics in Eq.~\eqref{eq:FOFHD_eq1}.

\subsection{Solutions of the saddle point equations \eqref{eq:saddle_olla}}
\label{app:sol-saddle-olla}
Following the saddle-point equations \eqref{eq:saddle_olla}, and assuming the derivatives of the saddle-point fields vanish at $X \to \pm \infty$, one finds (see Appendix~\ref{sec:delt_Wsd=0} for the derivation)
\begin{align}\label{eq:FO_delt_Wsd_0}
  \partial_t  \msc W^{\rm sd}_\lambda(v,t)=\partial_t\Big(\int dY\int du~\mf f^{\rm sd}_\lambda(Y,u,t)\mathcal{G}(u,v)\partial_Y\mf h^{\rm sd}_\lambda(Y,u,t)\Big)=0.
\end{align}
Hence $\msc W^{\rm sd}_\lambda(v,t)$ is independent of time. It is therefore convenient to rewrite the saddle-point equations in Eqs.~\eqref{eq:saddle_olla1}-\eqref{eq:saddle_olla2}, introducing the modified  effective velocity
\begin{align}\label{eq:FO_new eff_vel}
    \msc V^{\rm eff}_\lambda(v)=\msf V^{\rm eff}(v)+\msc W^{\rm sd}_\lambda(v),
\end{align}
as
\begin{subequations}
\begin{align}
  & \partial_t \mf f^{\rm sd}_\lambda(X,v,t)+\msc V^{\rm eff}_\lambda(v)\partial_X\mf f^{\rm sd}_\lambda(X,v,t)-\frac{\mc D(v)}{2}\partial^2_X\mf f^{\rm sd}_\lambda(X,v,t)=0\label{eq:new_saddle_olla1}\\
  & \partial_t\mf h^{\rm sd}_\lambda(X,v,t)+\msc V^{\rm eff}_\lambda(v)\partial_X\mf h^{\rm sd}_\lambda(X,v,t)+\frac{\mc D(v)}{2}\partial^2_X\mf h^{\rm sd}_\lambda(X,v,t)=0,\label{eq:new_saddle_olla2}
  \end{align}
\end{subequations}
while the Eqs.~\eqref{eq:saddle_olla3}-\eqref{eq:saddle_olla4} remain unchanged. The backward diffusion equation \eqref{eq:new_saddle_olla2}, together with the expression of $\mf h^{\rm sd}_\lambda(X,v,\msf T)$ in Eq.~\eqref{eq:saddle_olla3}, can be solved explicitly. This determines $\mf h^{\rm sd}_\lambda(X,v,0)$ as
\begin{align}\label{eq:FO_hxv0}
    \mf h^{\rm sd}_\lambda(X,v,0)=-\frac{\lambda}{2}~{\rm erfc}\Big(-\frac{X+\msc V^{\rm eff}_\lambda(v)\msf T}{\sqrt{2\mc D(v)\msf T}}\Big).
\end{align}
The initial phase-space density $\mf f^{\rm sd}_\lambda(X,v,0)$ at the saddle point can be determined from the saddle point equation Eq.~\eqref{eq:saddle_olla4}. Substituting the expression of the functional derivative of the free energy functional $\frac{\delta(\Delta \mathcal{F}^{\rm r}[\mf f(0)])}{\delta \mf f(X,v,0)}$ at the saddle point from Eq.~\eqref{eq:FOFHD_var_DelF_f} into Eq.~\eqref{eq:saddle_olla4}, we obtain
\begin{align}\label{eq:FO_fXv0}
    \mf f^{\rm sd}_\lambda(X,v,0)=\mf f_{\rm eq}(X,v)\Bigg(\frac{1-a\varrho^{\rm sd}_\lambda(X,0)}{1-a\varrho_{\rm eq}(X)}\Bigg)\exp\Bigg[\mathbb H_{\lambda}(X,v)-\frac{1}{1-a\varrho^{\rm sd}_\lambda(X,0)}+\frac{1}{1-a\varrho_{\rm eq}(X)}\Bigg],
\end{align}
where 
\begin{align}\label{eq:FO_fXv0_HXv_defn}
    \mathbb H_{\lambda}(X,v)=\lambda\Theta(X)+\mf h^{\rm sd}_\lambda(X,v,0)=\lambda\Theta(X)-\frac{\lambda}{2}~{\rm erfc}\Big(-\frac{X+\msc V^{\rm eff}_\lambda(v)\msf T}{\sqrt{2\mc D(v)\msf T}}\Big).
\end{align}
The saddle point density $\varrho^{\rm sd}_\lambda(X,0)$ is then given by integrating the expression in Eq.~\eqref{eq:FO_fXv0} over the velocity $v$ and can be expressed in terms of the Lambert $W$ function as
\begin{subequations}\label{eq:FO_rhoX0}
    \begin{align}
        \varrho^{\rm sd}_\lambda(X,0)=\frac{1}{a}\frac{W[\mc B_{\lambda}(X)]}{1+W[\mc B_{\lambda}(X)]},
    \end{align}
    with
    \begin{align}
        \mc B_{\lambda}(X)=\frac{a}{1-a\varrho_{\rm eq}(X)}\exp\Bigg(\frac{a\varrho_{\rm eq}(X)}{1-a\varrho_{\rm eq}(X)}\Bigg)\int dv~\mf f_{\rm eq}(X,v){\rm e}^{\mathbb H_{\lambda}(X,v)}.
    \end{align}
\end{subequations}
For a given $\msc V^{\rm eff}_\lambda(v)$, Eqs.~\eqref{eq:FO_hxv0}-\eqref{eq:FO_rhoX0} determine the initial saddle-point phase-space density $ \mf f^{\rm sd}_\lambda(X,v,0)$. However, according to Eq.~\eqref{eq:FO_new eff_vel}, the modified effective velocity $\msc V^{\rm eff}_\lambda(v)$ itself depends on the saddle-point field through $\msc W^{\rm sd}_\lambda(v)$, which in turn is determined by the saddle-point phase-space density. The saddle-point solution $ \mf f^{\rm sd}_\lambda(X,v,0)$ must therefore be determined self-consistently.
The solution of the forward diffusion equation \eqref{eq:new_saddle_olla1} together with  Eq.~\eqref{eq:FO_fXv0} is given by
\begin{align}\label{eq:FO_fXT}
    \mf f^{\rm sd}_\lambda(X,v,\msf T)=\int dY\frac{1}{\sqrt{2\pi\mc D(v) \msf T}}\exp[-\frac{\big(X-Y-\msc V^{\rm eff}_\lambda(v)\msf T\big)^2}{2\mc D(v)\msf T}]    \mf f^{\rm sd}_\lambda(Y,v,0).
\end{align}
Finally, evaluating Eq.~\eqref{eq:FO_fXT} at $X=0$, and using Eq.~\eqref{eq:saddle_olla5} together with Eq.  \eqref{eq:saddle_olla3}, we obtain the self-consistency relation
\begin{align}\label{eq:FO_Wsd_v}
   \msc W^{\rm sd}_\lambda(v)=-\lambda \int du~\mf f^{\rm sd}_\lambda(0,u,\msf T)\mc G(u,v),
\end{align}
where $\mf f^{\rm sd}_\lambda (0,u,\msf T)$ itself depends on $\msc V^{\rm eff}_\lambda$ through Eqs.~\eqref{eq:FO_hxv0} and \eqref{eq:FO_fXT}. The above equations therefore provide a self-consistent determination of the saddle-point phase-space density and the effective velocity. 
Having determined the saddle-point solutions from the saddle-point equations \eqref{eq:saddle_olla}, derived under the assumption that the action in Eq.~\eqref{eq:HR_FOFHD_action} scales linearly with time in the long-time limit, we now evaluate the saddle-point action \eqref{eq:S_saddle_Olla}. This allows us to verify a posteriori the consistency of the assumed long-time scaling.

\subsubsection{Derivation of $\partial_t  \msc W^{\rm sd}_\lambda(v,t)=0$ }\label{sec:delt_Wsd=0}
Following Eq.~\eqref{eq:saddle_olla5}, 
\begin{align}\label{eq:FO_delt_Wsd}
  \partial_t  \msc W^{\rm sd}_\lambda(v,t)&=\partial_t\Big(\int dY\int du~\mf f^{\rm sd}_\lambda(Y,u,t)\mathcal{G}(u,v)\partial_Y\mf h^{\rm sd}_\lambda(Y,u,t)\Big)\cr
 &=\int dY\int du~\mathcal{G}(u,v)\Big[\partial_Y\mf h^{\rm sd}_\lambda(Y,u,t)\partial_t\mf f^{\rm sd}_\lambda(Y,u,t)+\partial_Y\Big(\partial_t\mf h^{\rm sd}_\lambda(Y,u,t)\Big)\mf f^{\rm sd}_\lambda(Y,u,t)\Big]\cr
\end{align}
Substituting the expressions of $\partial_t\mf f^{\rm sd}_\lambda(Y,u,t),\partial_t\mf h^{\rm sd}_\lambda(Y,u,t)$ from Eqs.~\eqref{eq:saddle_olla1}-\eqref{eq:saddle_olla2} into Eq.~\eqref{eq:FO_delt_Wsd}, and simplifying, we obtain
\begin{align}\label{eq:FO_delt_Wsd_2}
    \partial_t  \msc W^{\rm sd}_\lambda(v,t)&= -\big(\msf V^{\rm eff}(v)+\msc W^{\rm sd}_\lambda(v,t)\big)\int du~\mathcal{G}(u,v)\int dY~\partial_Y\Big[\big(\partial_Y\mf h^{\rm sd}_\lambda(Y,u,t)\big)\mf f^{\rm sd}_\lambda(Y,u,t)\Big]\cr
    &~~~~-\frac{\mc D(v)}{2}\int du~\mathcal{G}(u,v)\int dY~\partial_Y\Big[\partial_Y^2\mf h^{\rm sd}_\lambda(Y,u,t)-\big(\partial_Y\mf h^{\rm sd}_\lambda(Y,u,t)\big)\big(\partial_Y\mf f^{\rm sd}_\lambda(Y,u,t)\big)\Big].\cr
\end{align}
Assuming the derivatives of the saddle-point fields $\mf f^{\rm sd}_\lambda(Y,u,t)$, and $\mf h^{\rm sd}_\lambda(Y,u,t)$ vanish at $Y \to \pm \infty$, both integrals over $Y$ vanish identically as total derivatives, yielding
\begin{align}
    \partial_t  \msc W^{\rm sd}_\lambda(v,t)=0.
\end{align}

\subsection{Consistency of the long-time scaling of the saddle-point action}
\label{app:action-large-T-FO}
To characterize its asymptotic behavior, we introduce the ballistic scaling variable $Z = X/\msf T$. At fixed $Z$, the argument of the complementary error function obtained in Eq.~\eqref{eq:FO_hxv0} grows as $\sqrt{\msf T}$. Therefore, in the limit $\msf T \to \infty$, the saddle-point field approaches the step-function form as
\begin{align}\label{Eq:FO_hXv0_largeT}
  \mt{h}^{\rm sd}_\lambda(Z,v,0)= \lim_{\msf T\to \infty} \mf h^{\rm sd}_\lambda(Z\msf T,v,0)=-\lambda\Theta(Z+\msc V^{\rm eff}_\lambda(v)),
\end{align}
depending only on the ballistic coordinate $Z$ and the effective velocity $-\msc V^{\rm eff}_\lambda(v)$. 
Consequently, following the definition of $\mathbb H_{\lambda}( X,v)$ given in Eq.~\eqref{eq:FO_fXv0_HXv_defn}, its large-time limit is
\begin{align}\label{eq:FO_saddle_scaled_H_lambda}
    \mt H_{\lambda}(Z,v)=\lim_{\msf T\to \infty}\mathbb H_{\lambda}(\msf TZ,v)=\lambda\big[\Theta(Z)-\Theta(Z+\msc V^{\rm eff}_\lambda(v))\big].
\end{align}
We next consider the corresponding large-time behavior of the saddle-point phase-space density. 
Under this ballistic scaling at large time, the saddle point phase space density, $\mt f^{\rm sd}_\lambda(Z,v,0)=\lim_{\msf T\to \infty}\mf f^{\rm sd}_\lambda(Z\msf T,v,0)$ follows from Eq.~\eqref{eq:FO_fXv0}, and takes the form
\begin{align}\label{eq:FO_fXv0_largeT}
    \mt f^{\rm sd}_\lambda(Z,v,0)=\mt f_{\rm eq}(Z,v)\Bigg(\frac{1-a\rho^{\rm sd}_\lambda(Z,0)}{1-a\rho_{\rm eq}(Z)}\Bigg)\exp\Bigg[ \mt H_{\lambda}(Z,v)-\frac{1}{1-a\rho^{\rm sd}_\lambda(Z,0)}+\frac{1}{1-a\rho_{\rm eq}(Z)}\Bigg],
\end{align}
where the quantities appearing in this expression are understood as their large-time limits. In particular, for the homogeneous equilibrium state considered here, we denote $\mt f_{\rm eq}(Z,v)=\lim_{\msf T\to \infty} \mf f_{\rm eq}(\msf TZ,v)$, $\rho_{\rm eq}(Z)=\lim_{\msf T\to \infty} \varrho_{\rm eq}(\msf TZ)$. The limiting saddle-point density denoted as $\rho^{\rm sd}_\lambda(Z,0)=\lim_{\msf T\to \infty} \varrho^{\rm sd}_\lambda(\msf TZ,0)$, is then obtained by taking the large-time limit of the Lambert $W$ expression given in Eq.~\eqref{eq:FO_rhoX0}. 

We next consider the solution of the forward diffusion equation, Eq.~\eqref{eq:FO_fXT}, in the large $\msf T$ limit. Introducing the ballistic variables $Z=X/\msf T,~R=Y/\msf T$, in the limit $\msf T\to \infty$, 
the Gaussian kernel following Eq.~\eqref{eq:FO_fXT}, $\sqrt{\frac{\msf T}{2\pi\mc D(v)}}\exp[-\frac{\msf T\big(Z-R-\msc V^{\rm eff}_\lambda(v)\big)^2}{2\mc D(v)}]$ becomes sharply localized around $R=Z-\msc V^{\rm eff}_\lambda(v)$ and converges to a Dirac delta distribution $\delta(Z-R-\msc V^{\rm eff}_\lambda(v))$.
Consequently, the large-time saddle-point phase-space density evaluates to 
\begin{align}\label{eq:FO_fXvT_largeT}
    \lim_{\msf T\to \infty} \mf f^{\rm sd}_\lambda(\msf TZ,v,\msf T)=\mt f^{\rm sd}_\lambda(Z-\msc V^{\rm eff}_\lambda(v),v,0).
\end{align}
For the cumulant generating function defined in Eq.~\eqref{eq:FO_CGF_Define}, using Eqs.~\eqref{eq:FO_fXv0_largeT} and \eqref{eq:FO_fXvT_largeT} in the expression of the integrated current in Eq. \eqref{def:Q(T)}, we find that the saddle-point current  $\mc Q(\msf T)|_{\mf f^{\rm sd}_\lambda,\mf h^{\rm sd}_\lambda}$ in the large $\msf T$ limit can be written as
\begin{align}
    \mc Q(\msf T)|_{\mf f^{\rm sd}_\lambda,\mf h^{\rm sd}_\lambda}=\msf T \msc Q[\mt f^{{\rm sd}},\mt h^{{\rm sd}}]
\end{align}
\begin{align}\label{eq:FO_saddle_Scaled_Q}
    {\rm with}~~\msc Q[\mt f^{{\rm sd}},\mt h^{{\rm sd}}]=\int dZ\int dv~\Theta(Z)[\mt f^{\rm sd}_\lambda(Z-\msc V^{\rm eff}_\lambda(v),v,0)-\mt f^{\rm sd}_\lambda(Z,v,0)]
\end{align}
The free energy functional in Eq.~\eqref{eq:S_saddle_Olla} has the same scaling in the large $\msf T$ limit. Indeed, introducing the scaling $X=\msf T Z$ and using the large-time form of the saddle-point phase-space density given in Eq.~\eqref{eq:FO_fXv0_largeT}, we obtain following Eq.~\eqref{eq:e^(-DelF)} as
\begin{align}
    \Delta\mathcal{F}^{\rm r}[\mf f^{\rm sd}_\lambda(0)]=\msf T \Delta\msc{F}^{\rm r}[\mt f^{{\rm sd}}(0)]
\end{align}
with
\begin{align}\label{eq:FO_LargeT_F}
    \Delta\msc{F}^{\rm r}[\mt f^{{\rm sd}}(0)]=\int dZ\int dv\Bigg[\mt f^{\rm sd}_\lambda(Z,v,0)\left(\mt H_{\lambda}(Z,v)-\frac{1}{1-a\rho^{\rm sd}_\lambda(Z,0)}\right)+\frac{\mt f_{\rm eq}(Z,v)}{1-a\rho_{\rm eq}(Z)}\Bigg],
\end{align}
where recall $\mt f^{\rm sd}_\lambda(Z,v,0)$ is given in Eq.~\eqref{eq:FO_fXv0_largeT}. Finally, we consider the  contribution of the third term  $\frac{1}{2}\int_0^{\msf T}dt\int dX\int dv ~\mf f^{\rm sd}_\lambda(X,v,t)~\partial_X \mf h^{\rm sd}_\lambda(X,v,t) \msc W^{\rm sd}_\lambda(v,t)$ to the saddle-point action in Eq.~\eqref{eq:S_saddle_Olla}. As shown in Eq.~\eqref{eq:FO_delt_Wsd_0}, $\msc W^{\rm sd}_\lambda(v)$ is time-independent. Moreover, using the saddle-point equations \eqref{eq:saddle_olla} and assuming that the derivatives of the saddle-point fields vanish at spatial infinity, one finds 
\begin{align}
    \partial_t\Big(\int dX~\mf f^{\rm sd}_\lambda(X,v,t)~\partial_X \mf h^{\rm sd}_\lambda(X,v,t)\Big)=0.
\end{align}
Hence the spatial integral is independent of time, and the contribution from this term is exactly proportional to $\msf T$. We have thus shown that, in the large-$\msf T$ limit, each of the three contributions to the saddle-point action in Eq.~\eqref{eq:S_saddle_Olla} is proportional to $\msf T$. Consequently, the cumulant generating function in Eq.~\eqref{eq:FO_CGF_Define}, at large $\msf T$ limit, takes the form 
\begin{align}\label{eq:FO_Gamma_scaleT}
   \msc G_{\rm F}(\lambda,\msf T)=\msf T \Gamma_{\rm cb}^{\rm r}(\lambda),
\end{align}
where the explicit expression of $\Gamma_{\rm cb}^{\rm r}(\lambda)$ is obtained by evaluating each of the three contributions to the saddle-point action in Eq.~\eqref{eq:S_saddle_Olla}, as we now discuss in turn. Following Eq.~\eqref{eq:FO_saddle_Scaled_Q} the current contribution $\lambda \msc Q[\mt f^{{\rm sd}},\mt h^{{\rm sd}}]$ can be written as
\begin{align}
    \lambda \msc Q[\mt f^{{\rm sd}},\mt h^{{\rm sd}}]=-\int dZ\int dv~\mt H_{\lambda}(Z,v)\mt f^{\rm sd}_\lambda(Z,v,0)
\end{align}
where $\mt H_{\lambda}(Z,v)$ is given in Eq.~\eqref{eq:FO_saddle_scaled_H_lambda}. Following Eq.~\eqref{eq:FO_LargeT_F}, we therefore write
\begin{align}
   \Delta\msc{F}^{\rm r}[\mt f^{{\rm sd}}(0)]+\lambda \msc Q[\mt f^{{\rm sd}},\mt h^{{\rm sd}}]= \int dZ\Big[\frac{\rho_{\rm eq}(Z)}{1-a\rho_{\rm eq}(Z)}-\frac{\rho^{\rm sd}_\lambda(Z,0)}{1-a\rho^{\rm sd}_\lambda(Z,0)}\Big],
\end{align}
where, the large-time saddle-point density $\rho^{\rm sd}_\lambda(Z,0)$ is obtained from the large-time limit of Eq.~\eqref{eq:FO_rhoX0} as
\begin{subequations}\label{eq:FO_rhoX0_scaled}
    \begin{align}
        \rho^{\rm sd}_\lambda(Z,0)=\frac{1}{a}\frac{W[\msc B_{\lambda}(Z)]}{1+W[\msc B_{\lambda}(Z)]},
    \end{align}
    with
    \begin{align}
        \msc B_{\lambda}(Z)=\frac{a}{1-a\rho_{\rm eq}(Z)}\exp\Bigg(\frac{a\rho_{\rm eq}(Z)}{1-a\rho_{\rm eq}(Z)}\Bigg)\int dv~\mt f_{\rm eq}(Z,v){\rm e}^{\mt H_{\lambda}(Z,v)}.
    \end{align}
\end{subequations}
We now  focus on obtaining the large $\msf T$ asymptotic of the third term in the saddle-point action in detail.  Since $\msc W^{\rm sd}_\lambda(v)$ is independent of time, and the spatial integral $\int dX~\mf f^{\rm sd}_\lambda(X,v,t)\partial_X \mf h^{\rm sd}_\lambda(X,v,t)$ is likewise time-independent (see discussion above), the third term in Eq.~\eqref{eq:S_saddle_Olla} is proportional to $\msf T$. To obtain the proportionality factor, it therefore suffices to evaluate this contribution at $t=0$, {\it i.e.} to compute $\int dX\int dv ~\mf f^{\rm sd}_\lambda(X,v,0)~\partial_X \mf h^{\rm sd}_\lambda(X,v,0) \msc W^{\rm sd}_\lambda(v)$. 
From Eq.~\eqref{eq:FO_hxv0}, the spatial derivative $\partial_X\mf h^{\rm sd}_\lambda(X,v,0)$
is a Gaussian centered at $X=-\msc V^{\rm eff}_\lambda(v)\msf T$ with width of order $\sqrt{\mathcal D(v)/\msf T}$. Although $\mf h^{\rm sd}_\lambda$ itself converges to a step function in the large time limit, as in Eq.~\eqref{Eq:FO_hXv0_largeT}, its spatial derivative retains a nontrivial contribution from the boundary layer around the discontinuity. Consequently, the limiting step-function form cannot be differentiated directly to evaluate this term. One must therefore retains the full expression for $\mf h^{\rm sd}_\lambda(X,v,0)$ given in Eq.~\eqref{eq:FO_hxv0} while evaluating the spatial integral $\int dX~\mf f^{\rm sd}_\lambda(X,v,0)\partial_X \mf h^{\rm sd}_\lambda(X,v,0)$. In the large time limit, this gives 
\begin{subequations}\label{eq:FO_larget_F_function}
    \begin{align}
    \lim_{\msf T\to \infty}\int dX~\mf f^{\rm sd}_\lambda(X,v,0)\partial_X \mf h^{\rm sd}_\lambda(X,v,0)=-\lambda F(-\msc V^{\rm eff}_\lambda(v),v)
\end{align}
where
\begin{align}\label{eq:FO_larget_F_function2}
    F\big(-\msc V^{\rm eff}_\lambda(v),v\big)=&\mt f_{\rm eq}\big(-\msc V^{\rm eff}_\lambda(v),v\big)~\frac{1-a\rho^{\rm sd}_\lambda(-\msc V^{\rm eff}_\lambda(v))}{1-a\rho_{\rm eq}(-\msc V^{\rm eff}_\lambda(v))}\pa{\frac{1-e^{-\lambda}}{\lambda}}\\
    &\exp\pa{-\frac{1}{1-a\rho^{\rm sd}_\lambda(-\msc V^{\rm eff}_\lambda(v))}+\frac{1}{1-a\rho_{\rm eq}(-\msc V^{\rm eff}_\lambda(v))}+\lambda\Theta(-\msc V^{\rm eff}_\lambda(v))}.\notag
\end{align}
\end{subequations}
Substituting the result of Eq.~\eqref{eq:FO_larget_F_function} into the definition of $\msc W^{\rm sd}_\lambda(v)$ in Eq.~\eqref{eq:saddle_olla5}, we find that its large time limit $\msf{W}^{\rm sd}_\lambda(v)=\lim_{\msf T\to \infty}\msc W^{\rm sd}_\lambda(v)$ satisfies the self-consistency relation
\begin{align}\label{eq:FO_larget_W_sd}
    \msf{W}^{\rm sd}_\lambda(v)=-\lambda\int du~F(-\msc V^{\rm eff}_\lambda(u),u)\mc G(u,v).
\end{align} 
together with the expression of $\msc V^{\rm eff}_\lambda(u)$ in Eq.~\eqref{eq:FO_new eff_vel}. Collecting all three contributions in the large $\msf T$ limit, the scaled cumulant generating function $\Gamma_{\rm cb}^{\rm r}(\lambda)$ takes the explicit form as in Eq.~\eqref{eq:FO_gamma_lambda-mt}.

\subsection{Derivation of Eq.~\eqref{eq:FO_Q2_expression}}\label{sec:2nd_cum_FOFHD}
To determine the second cumulant of the integrated current, we expand the saddle-point phase-space density $\mt f^{\rm sd}_\lambda(Z,v,0)$ and the corresponding mass density $\rho^{\rm sd}_\lambda(Z,0)$  in powers of $\lambda$ as
\begin{align}\label{eq:FO_f_lambda}
    \mt f^{\rm sd}_\lambda(Z,v,0)=\mt f_{\rm eq}(Z,v)+\lambda \mt f_{1}(Z,v,0) + \lambda^2 \mt f_{2}(Z,v,0)
\end{align}
\begin{align}\label{eq:FO_rho_lambda}
    \rho^{\rm sd}_\lambda(Z,0)=\rho_{\rm eq}(Z)+\lambda \rho_{1}(Z,0) + \lambda^2 \rho_{2}(Z,0).
\end{align}
From Eq.~\eqref{eq:FO_gamma_lambda-mt}, the term in the rhs, $\frac{\lambda^2}{2}\int dv\int du~F(-\msc V^{\rm eff}_\lambda(v),v)F(-\msc V^{\rm eff}_\lambda(u),u)\mc G(u,v)$ contributes only from $\mathcal{O}(\lambda^2)$ onward. At zeroth order in $\lambda$, the saddle-point solution coincides with the equilibrium solution, $\mt f^{\rm sd}_\lambda=\mt f_{\rm eq}$ and $\rho^{\rm sd}_\lambda=\rho_{\rm eq}$. Hence, the zeroth-order contribution to $\Gamma_{\rm cb}^{\rm r}(\lambda)$ in Eq.~\eqref{eq:FO_gamma_lambda-mt} vanishes identically. We first consider the terms linear in $\lambda$ to determine the mean integrated current. Since $\msf{W}^{\rm sd}_\lambda(v)$ in Eq.~\eqref{eq:FO_larget_W_sd} is proportional to $\lambda$, the effective velocity in Eq.~\eqref{eq:FO_new eff_vel} admits the expansion
\begin{align}\label{eq:FO_Veff_Expand}
 \msc V^{\rm eff}_\lambda(v) =  \msf V^{\rm eff}(v)+\lambda\mt W_1(v)+\mathcal{O}(\lambda^2), ~{\rm with}~\mt W_1(v)=-\int du~\mt f_{\rm eq}(-\msc V^{\rm eff}_\lambda(u),u)\mc G(u,v).
\end{align}
Substituting this into Eq.~\eqref{eq:FO_saddle_scaled_H_lambda}, the function $\mt H_\lambda(Z,v)$ expands as
\begin{align}\label{eq:FO_H_expand_lambda}
    \mt H_{\lambda}(Z,v)=-\lambda \mt H_0(Z,v)-\lambda^2\delta\big(Z+\msf V^{\rm eff}(v)\big) \mt W_1(v)+\mathcal{O}(\lambda^3),
\end{align}
where
\begin{align}
    \mt H_0(Z,v)=\Theta(Z+\msf V^{\rm eff}(v))-\Theta(Z).
\end{align}
To expand the saddle-point density via the Lambert-$W$ expression in Eq.~\eqref{eq:FO_rhoX0_scaled}, it is convenient to first expand the quantity
$\mathcal A_{\lambda}(Z)=\int dv~\mt f_{\rm eq}(Z,v){\rm e}^{\mt H_{\lambda}(Z,v)}$, which using Eq.~\eqref{eq:FO_H_expand_lambda} can be written as
\begin{subequations}\label{eq:FO_A_expand}
    \begin{align}
        \mathcal A_{\lambda}(Z)=\mathcal A_0(Z)+\lambda \mathcal A_1(Z)+\lambda^2\mathcal A_2(Z)
    \end{align}
    with the coefficients
    \begin{align}\label{eq:FO_A_expand_1}
        &\mathcal A_0(Z)=\rho_{\rm eq}(Z),~\mathcal A_1(Z)=-\int dv~ \mt f_{\rm eq}(Z,v)~\mt H_0(Z,v)\cr
        &\mathcal A_2(Z)=\frac{1}{2}\int dv~ \mt f_{\rm eq}(Z,v)~\mt H_0^2(Z,v)-\int dv~ \mt f_{\rm eq}(Z,v) ~\delta(Z+\msf V^{\rm eff}(v))\mathcal{W}_1(v).
    \end{align}
\end{subequations}
Expanding the Lambert $W$ expression for the saddle-point density $\rho^{\rm sd}_\lambda(Z,0)$ in Eq.~\eqref{eq:FO_rhoX0_scaled}, using Eq.~\eqref{eq:FO_A_expand}, up to second order in $\lambda$, and incorporating the $\mathcal{O}(\lambda^2)$ contribution from the last term in Eq.~\eqref{eq:FO_gamma_lambda-mt} the cumulant generating function $\Gamma(\lambda)$ up to $\mathcal{O}(\lambda^2)$ reads
\begin{align}\label{eq:FO_CGF_lambda}
    \Gamma(\lambda)=&\lambda \int dZ \mathcal{I}(Z)-\lambda^2\int dZ\Big[\mathcal A_2(Z)-\frac{a}{2}\big(2-a\rho_{\rm eq}(Z)\big)\mathcal I^2(Z)\Big]\cr
    &~~~~~+\frac{\lambda^2}{2} \int du\int dv~\mt f_{\rm eq}(-\msc V^{\rm eff}_\lambda(u),u)\mt f_{eq}(-\msc V^{\rm eff}_\lambda(v),v)\mc G(u,v)+\mathcal{O}(\lambda^3) 
\end{align}
where
\begin{align}
    \mathcal I(Z)=\int dv~\mt f_{\rm eq}(Z,v)\big[\Theta(Z+\msf V^{\rm eff}(v))-\Theta(Z)\big]=-\mathcal{A}_1(Z)
\end{align}
and $\mathcal A_2(Z)$ is given in Eq.~\eqref{eq:FO_A_expand_1}. The term linear in $\lambda$ determines the mean integrated current, while the coefficient of the quadratic term determines the second cumulant. Using the expression of $\mt W_1(v)$ from Eq.~\eqref{eq:FO_Veff_Expand} in \eqref{eq:FO_A_expand_1} and inserting the resulting expression for $\mathcal A_2(Z)$ into Eq.~\eqref{eq:FO_CGF_lambda}, we obtain the second cumulant as
\begin{align}
  \langle \mc Q(\msf T)^2 \rangle_{c,{\rm eq}}^{\rm cb}=&\msf T\Bigg[\int dZ\int dv~\mt f_{\rm eq}(Z,v)~\mt H_0^2(Z,v)-\int dZ~a\big(2-a\rho_{\rm eq}(Z)\big) \mathcal I^2(Z)\cr
  &~~~+ \int du\int dv~\mt f_{\rm eq}(-\msc V^{\rm eff}_\lambda(u),u)\mt f_{eq}(-\msc V^{\rm eff}_\lambda(v),v)\mc G(u,v)\Bigg],
\end{align}
where recall $\mt H_0(Z,v)=\Theta(Z+\msf V^{\rm eff}(v))-\Theta(Z)$, and $\mc G(u,v)$ is defined in Eq.~\eqref{eq:G_uv}.
For the homogeneous equilibrium state, considered in this paper, $\mt f_{\rm eq}(Z,v)=\varrho_0\mc g(v)$, where $\mc g(v)$ is the zero-mean Maxwell distribution at temperature $T_0$, and $\rho_{\rm eq}(Z)=\varrho_0$, we get $\int dZ~\mathcal I(Z)=0$, confirming that the mean integrated current vanishes. The explicit expression of the variance becomes
\begin{align}\label{eq:FO_Q2_expression_supp}
    \langle \mc Q(\msf T)^2 \rangle_{c,{\rm eq}}^{\rm cb}=\msf T\frac{\varrho_0\sqrt{T_0}}{\sqrt{\pi}(1-a\varrho_0)}\Big[\sqrt{2}-(2a-a^2\varrho_0)\varrho_0(\sqrt{2}-1)+ a^2\rho_0^2\Big].
\end{align}

\bibliographystyle{unsrt}
\bibliography{references}
\end{document}

%% file: macro.tex
\newcommand{\pa}[1]{\left( #1 \right)}
